\documentclass[aps,preprint,nofootinbib,showpacs]{revtex4}
\usepackage{graphicx,color,amsmath}
\newcommand{\be}{\begin{equation}}
\newcommand{\ee}{\end{equation}}
\newcommand{\ba}{\begin{eqnarray}}
\newcommand{\beq}{\begin{equation}}
\newcommand{\eeq}{\end{equation}}
\newcommand{\ea}{\end{eqnarray}}

\def\beqa{\begin{eqnarray}}
\def\eeqa{\end{eqnarray}}

\def\bea{\begin{eqnarray}}
\def\eea{\end{eqnarray}}

\def\err#1#2{\lower2pt\hbox{ $\stackrel{\scriptstyle +#1}{\scriptstyle -#2}$}}
\def\ga{\mathrel{\raise.3ex\hbox{$>$\kern-.75em\lower1ex\hbox{$\sim$}}}}
\def\la{\mathrel{\raise.3ex\hbox{$<$\kern-.75em\lower1ex\hbox{$\sim$}}}}
\def\bmaT{\left(\begin{array}{ccc}}
\def\emaT{\end{array}\right)}
\def\bma{\left( \begin{array} }
\def\ema{\end{array} \right)}
\def\gsim{~{\rlap{\lower 3.5pt\hbox{$\mathchar\sim$}}\raise 1pt\hbox{$>$}}\,}
\def\lsim{~{\rlap{\lower 3.5pt\hbox{$\mathchar\sim$}}\raise 1pt\hbox{$<$}}\,}

\begin{document}

\preprint{
\vbox{%
\hbox{SHEP-12-01}
\hbox{DCP-12-01}
}}
\title{\boldmath Light charged Higgs bosons decaying to charm and bottom \\
quarks in models with two or more Higgs doublets \unboldmath} 
\author{A.G. Akeroyd}
\email{a.g.akeroyd@soton.ac.uk}
\author{S. Moretti}
\email{s.moretti@soton.ac.uk}
\affiliation{School of Physics and Astronomy, University of Southampton, \\
Highfield, Southampton SO17 1BJ, United Kingdom,}
\affiliation{Particle Physics Department, Rutherford Appleton Laboratory, Chilton, Didcot, 
Oxon OX11 0QX, United Kingdom}
\author{J. Hern\'andez-S\'anchez}
\affiliation{Facultad de Ciencias de la Electr\'onica, Benem\'erita Universidad Aut\'onoma de
Puebla, Apdo. Postal 542, 72570 Puebla, Puebla, M\'exico,
and Dual C-P Institute of High Energy Physics, M\'exico}
\email{jaimeh@ece.buap.mx}

%\date{\today}
\begin{abstract}
\noindent
Searches for light charged Higgs bosons ($H^\pm$) in the decay of top quarks,
$t\to H^\pm b$, are being carried out at the LHC and at the Tevatron. It is assumed that the dominant
decay channels for such an $H^\pm$ state are either $H^\pm\to \tau\nu$ or $H^\pm\to cs$, and separate searches 
are performed with comparable sensitivity to the parameters $m_{H^\pm}$ and $\tan\beta$ of the scalar potential.  The branching ratio for the 
decay $H^\pm\to cb$ can be as large as $80\%$ 
in the Aligned Two Higgs Doublet Model and in models with three or more Higgs doublets 
with natural flavour conservation, while satisfying the constraint 
from $b\to s\gamma$ for $m_{H^\pm}< m_t$. Although the current search strategy for
$H^\pm\to cs$ is also sensitive to $H^\pm\to cb$, a considerable gain in 
sensitivity could be obtained by tagging the $b$ quark from the decay $H^\pm\to cb$. Such an analysis, which could
be readily performed at the Tevatron and in  the 7 TeV and 8  TeV runs of the LHC, would
probe a parameter space of the fermionic couplings of $H^\pm$ in 
the above models which at present cannot be probed by experimental observables in flavour physics.

\end{abstract}
%\keywords: Higgs boson, Neutrino mass and mixing
\maketitle
%%%%%%%%%%%%%%%%%%%%%%%%%%%%%%%%%%%%%%%%%%%%%%%%%%

%%%%%%%%%%%%%%%%%%%%%%%%%%%%%%%%%%%%%%%%%%%%%%%%%%%%%%%%%%%%%%%%%%%%%%

\section{Introduction}

There is much ongoing experimental effort by the ATLAS and CMS collaborations at the CERN 
Large Hadron Collider (LHC)
to search for the neutral Higgs boson ($h^0$) of the Standard Model (SM) \cite{ATLAS:2012ae,Chatrchyan:2012tx}.
This model of spontaneous symmetry breaking  will be tested at the 
LHC over all of the theoretically preferred mass range,
in an experimental programme which is expected to be completed by the end of the 8 TeV run of the LHC.
At present \cite{latest_LHC} there are only two regions for the mass for $h^0$ which have not been excluded at 95\% c.l:
 i) a region of light mass, with $122 \,{\rm GeV} < m_{h^0} \,< 128 \,{\rm GeV}$, and ii) a region
of heavy mass, $m_{h^0}> 600$ GeV.

However, this simplest model of one fundamental scalar with a 
vacuum expectation value (vev) might not be nature's choice. There could be
additional scalar fields which also contribute to the masses of the fermions and weak bosons, with a 
more complicated scalar potential which depends on several arbitrary parameters.  
Importantly, even in the event of no signal for a SM-like Higgs boson at the LHC, the search for
scalar particles should continue in earnest due to the fact that a non-minimal Higgs sector
can give rise to different experimental signatures, some of which are challenging to detect.
Consequently, it will take much longer for the LHC to probe all of the parameter space of such models.

A commonly-studied extension of the Higgs sector of the SM is the ``Two Higgs Doublet Model'' (2HDM), 
which is composed of two Higgs isospin doublets \cite{Lee:1973iz} (this model has recently been reviewed   
in \cite{Branco:2011iw}). Notably, this structure is necessary in the Minimal Supersymmetric 
(SUSY) extension of the SM (called the ``MSSM''). The extra Higgs doublet gives rise to a
particle spectrum with multiple Higgs bosons; three are electrically neutral (two are CP-even, one is CP-odd) and two
are electrically charged (denoted by $H^\pm$). Flavour-changing neutral currents (FCNCs) mediated by scalars at tree level can be eliminated
by requiring that the scalar interactions with the fermions are invariant under discrete symmetries 
(``natural flavour conservation'', NFC) \cite{Glashow:1976nt}. 
The discovery of a charged scalar $H^\pm$ would be unequivocal evidence of a non-minimal Higgs sector, and
there have been many studies of the prospects of directly observing 
$H^\pm$ from a 2HDM or the MSSM at the Tevatron and LHC \cite{Barger:1989fj} (for reviews see \cite{Branco:2011iw,Roy:2004az})
Moreover, the effect of $H^\pm$ on the decay rates of mesons (especially $B$ mesons) also plays a major role
in constraining $m_{H^\pm}$ and the fermionic couplings of $H^\pm$ 
\cite{Barger:1989fj},\cite{Eriksson:2008cx}.

If $m_{H^\pm} <  m_t+m_b$, such particles would most copiously 
(though not exclusively  \cite{Aoki:2011wd}) be produced in the decays of top quarks via 
$t\to H^\pm b$ \cite{tbH}.
Searches in this channel have been performed by the Tevatron experiments, assuming 
the decay modes $H^\pm\to cs$ and  $H^\pm\to \tau\nu$ \cite{:2009zh,Aaltonen:2009ke}. 
Since no signal has been observed, constraints are obtained on the parameter space of $[m_{H^\pm}, \tan\beta]$, 
where $\tan\beta=v_2/v_1$ (i.e. the ratio of the vacuum expectation values of the two scalar doublets). 
Searches in these channels have now been carried out at the LHC:
i) for $H^\pm\to cs$ with 0.035 fb$^{-1}$ by ATLAS \cite{ATLAS:search}, and ii) 
for $H^\pm\to \tau\nu$ with 4.8 fb$^{-1}$ by ATLAS \cite{ATLAS_Htau} and with 1 fb$^{-1}$ by CMS \cite{CMS_Htau}.
These are the first searches for $H^\pm$ at this collider. The constraints on $[m_{H^\pm}, \tan\beta]$ 
from the LHC searches for $t\to H^\pm b$
are now superior to those obtained from the corresponding Tevatron searches.
 
The phenomenology of $H^\pm$ in models with three or more Higgs doublets  
(called Multi-Higgs Doublet Models, MHDM), was first studied comprehensively in \cite{Grossman:1994jb}, with an
emphasis on the constraints from low-energy processes (e.g. the decays of mesons).
Although the phenomenology of $H^\pm$ at high-energy colliders in a MHDM and in a 2HDM has many similarities,
the possibility of $m_{H^\pm}< m_t$ together with an enhanced branching ratio (BR) for 
$H^\pm\to cb$ would be a distinctive feature of the MHDM. This scenario, which was
first mentioned in \cite{Grossman:1994jb} and studied in more detail in 
\cite{Akeroyd:1994ga,Akeroyd:1995cf,Akeroyd:1998dt}, is of immediate interest for the
ongoing searches for $t\to H^\pm b$ with $H^\pm\to cs$ by the LHC \cite{ATLAS:search}.
Although the current limits on $H^\pm\to cs$  can be applied to
the decay $H^\pm \to cb$ (as discussed in \cite{Logan:2010ag} in the context of the Tevatron searches), a further
improvement in sensitivity to  $t\to H^\pm b$ with $H^\pm \to cb$ could be obtained by tagging the 
$b$ quark which originates from $H^\pm$ \cite{Akeroyd:1995cf,DiazCruz:2009ek,Logan:2010ag}. 
We will estimate the increase in sensitivity to BR$(H^\pm \to cb)$ and 
to the fermionic couplings of
$H^\pm$ in this scenario. 

Large values of BR$(H^\pm \to cb)$ are also possible in certain 2HDMs,
such as  the ``flipped 2HDM''  with 
NFC \cite{Akeroyd:1994ga,Aoki:2009ha,Logan:2010ag}.
However, in this model one would expect $m_{H^\pm} > m_t$ due to the 
constraint from $b\to s \gamma$  ($m_{H^\pm} > 295$ GeV \cite{Hou:1987kf,Borzumati:1998tg,Misiak:2006zs}), and thus
$t\to H^\pm b$ with $H^\pm \to cb$ would not proceed unless there were additional 
New Physics beyond that of the 2HDM which contributed to $b\to s \gamma$, and weakened the constraint on $m_{H^\pm}$. In the 
"Aligned Two Higgs Doublet Model" (A2HDM) \cite{Pich:2009sp} 
there are no FCNCs (as is the case in a 2HDM with NFC) 
due to an alignment of the Yukawa couplings.
The phenomenology of 
$H^\pm$ in the A2HDM \cite{Jung:2010ik} is very similar to that of $H^\pm$ in a MHDM
\cite{Cree:2011uy} , and $m_{H^\pm} < m_t$ in the A2HDM is also compatible with constraints from
$b\to s\gamma$. Our numerical results for BR($H^\pm \to cb$) in a MHDM apply directly to the A2HDM. 
The 2HDM without NFC and without alignment
also has a sizeable parameter space for a large BR$(H^\pm \to cb)$, and a detailed study can be
found in \cite{DiazCruz:2009ek}.

Our work is organised as follows. In section II we describe the fermionic interactions of 
$H^\pm$ in the MHDM/A2HDM. In section III we quantify the parameter space for 
a large BR$(H^\pm\to cb)$ in the MHDM/A2HDM. In section IV we summarise the
Tevatron/LHC searches for $t\to H^\pm b$ with $H^\pm\to cs$ and discuss how they
could be optimised for $H^\pm\to cb$. Section V contains our numerical results, 
with conclusions in section VI.

\section{$H^\pm$ in models with more than two Higgs doublets and in the Aligned 2HDM}

In a general 2HDM each fermion type (i.e. up-type quarks, down-type quarks and charged leptons)
couples to both of the scalar doublets. This would lead to FCNCs which are mediated
by the neutral scalars, and the magnitude of the associated Yukawa coupling is constrained
by experimental data (especially meson-antimeson mixing and  the decays of mesons). Such FCNCs can be suppressed by assuming that 
the flavour-changing Yukawa couplings are very small, which can be achieved by invoking a 
specific structure of the
fermion mass matrices \cite{Cheng:1987rs}. An alternative approach is to eliminate the FCNCs by requiring that 
the Lagrangian is invariant under a discrete symmetry, 
which is achieved if each species of fermion couples to at most one scalar doublet (NFC).
This condition leads to four distinct types of 2HDMs which differ in their Yukawa couplings.
These four models are called Model I, Model II, Lepton-specific 
and Flipped.\footnote{The Lepton-specific and Flipped models are referred to as IIA and IIB in \cite{Barnett:1983mm},
IV and III in \cite{Barger:1989fj}, 
I$'$ and II$'$ in \cite{Grossman:1994jb,Akeroyd:1994ga,Akeroyd:1995cf,Akeroyd:1998dt} and
$X$ and $Y$ in \cite{Aoki:2009ha}.} Models I and II
have received much phenomenological attention, while the study of the Lepton-specific and Flipped models
has been revived recently \cite{Barger:2009me,Aoki:2009ha,Su:2009fz,Logan:2010ag}, with early studies in 
\cite{Akeroyd:1994ga,Akeroyd:1998dt,Akeroyd:1996di}. We now introduce the fermionic couplings of
$H^\pm$ in a MHDM and the A2HDM, and discuss how these couplings differ from those in the above 2HDMs with NFC.

\subsection{The Multi-Higgs Doublet Model (MHDM)}
A MHDM is an extension of the 2HDM 
with $n$ scalar doublets, where $n \geq 3$. The suppression of FCNCs is obtained by imposing NFC.
As in the 2HDM, the MHDM has the virtue of predicting $\rho=1$ at tree level,
with finite higher-order corrections which depend on the mass splittings of the scalars. 
In the MHDM there are $n-1$ charged scalars, and a detailed study of the phenomenology of 
the lightest $H^\pm$ in such models was performed in \cite{Grossman:1994jb},
with the assumption that the other $H^\pm$ are much heavier.
The interaction of the lightest $H^\pm$ in a MHDM with the fermions
is described by the following Lagrangian:
\begin{equation}
{\cal L}_{H^\pm} =
-\left\{\frac{\sqrt2V_{ud}}{v}\overline{u}
\left(m_d X{P}_R+m_u Y{P}_L\right)d\,H^+
+\frac{\sqrt2m_e }{v} Z\overline{\nu_L^{}}\ell_R^{}H^+
+{H.c.}\right\}
\label{lagrangian}
\end{equation}

Here $u$ and $d$ denote up-type quarks and down-type quarks respectively (for all three generations); 
$V_{ud}$ is a CKM matrix element; $m_u$, $m_d$ and $m_e$ are the masses of the up-type quarks, down-type quarks and charged leptons
respectively; $P_L$ and $P_R$ are chirality projection operators, and $v=246$ GeV. In a 2HDM with natural flavour conservation, 
the couplings $X$, $Y$ and $Z$ are determined solely by $\tan\beta=v_2/v_1$. 
The values of $X$, $Y$ and $Z$ in the four versions of the 2HDM \cite{Barger:1989fj} 
are given in Table~\ref{couplings}.
It is clear that $|X|,|Y|$ and $|Z|$ are simply related in the 2HDM e.g. 
one has  $|X|=|Z|=1/|Y|$ for the Type II structure.

\begin{table}[h]
\begin{center}
\begin{tabular}{|c||c|c|c|}
\hline
& $X$ &  $Y$ &  $Z$ \\ \hline
Type I
&  $-\cot\beta$ & $\cot\beta$ & $-\cot\beta$ \\
Type II
& $\tan\beta$ & $\cot\beta$ & $\tan\beta$ \\
Lepton-specific
& $-\cot\beta$ & $\cot\beta$ & $\tan\beta$ \\
Flipped
& $\tan\beta$ & $\cot\beta$ & $-\cot\beta$ \\
\hline
\end{tabular}
\end{center}
\caption{The couplings $X$,$Y$ and $Z$ in the Yukawa interactions of $H^\pm$ in the four versions of the 2HDM with natural flavour
conservation.}
\label{couplings}
\end{table}

In a MHDM the couplings $X$, $Y$ and $Z$ are {\it arbitrary complex numbers}, which are defined in terms of the
$n\times n$ matrix $U$ which  diagonalises the mass matrix of the charged scalars:
\begin{equation}
X_i=\frac{U_{di}}{U_{d1}}, \;\,
Y_i=-\frac{U_{ui}}{U_{u1}}, \;\,
Z_i=\frac{U_{ei}}{U_{e1}}.
\end{equation}
We follow the notation of \cite{Grossman:1994jb}
in which $i=1$ corresponds to the couplings of the charged Goldstone boson, 
and $i$ runs from 2 to $n$ for the physical charged scalars. The fermionic couplings of the lightest $H^\pm$ 
in a MHDM are taken to be $X_2$, $Y_2$ and $Z_2$.
The subscripts $d,u,e$ take any integer value up to $n$, 
and specify which of the $n$ doublets couples to which fermion type
e.g. for a Type II structure one sets $d=e=1$ and $u=2$, while for the ``democratic''
3HDM (e.g. \cite{Cree:2011uy}) one has $d=1$, $u=2$ and $e=3$.
In a 2HDM, $U$ is a $2\times 2$ matrix with elements given by $\sin\beta$ 
and $\cos\beta$ (i.e. one free parameter). In a 3HDM, $U$ is a $3\times 3$ matrix with
four free parameters which can be taken as $\tan\beta=v_u/v_d$, $\tan\gamma=\sqrt{(v_d^2+v_u^2)}/v_e$,
a mixing angle $\theta$ for the two $H^\pm$, and a complex phase $\delta$. An explicit form of the
matrix $U$ for the 3HDM is given in \cite{Cree:2011uy}.

Due to the unitarity of the matrix $U$ one can derive the following identities \cite{Grossman:1994jb}:
\begin{equation}
\sum_{i=2}^n X_iY_i^*=1 \,\,{\rm (for} \, d\not=u),
\label{sumruleXY}
\end{equation}
\begin{equation}
\sum_{i=2}^n X_iZ_i^*=-1 \,\,{\rm (for} \, d\not=e), 
\label{sumruleXZ}
\end{equation}
\begin{equation}
\sum_{i=2}^n Y_iZ_i^*=1 \,\,{\rm (for} \, u\not=e), 
\label{sumruleYZ}
\end{equation}
and
\begin{equation}
\sum_{i=2}^n {|X_i|}^2= \frac{v^2}{v_d^2}-1, \;
\sum_{i=2}^n {|Y_i|}^2= \frac{v^2}{v_u^2}-1, \;
\sum_{i=2}^n {|Z_i|}^2= \frac{v^2}{v_e^2}-1.
\label{moresumrules}
\end{equation}

In a 2HDM these identities reduce to simple trigonometric relations involving $\tan\beta$. 
It is evident that the branching ratios of
$H^\pm$ to fermions in the MHDM depend on the three-dimensional parameter space of $X_i$, $Y_i$ and $Z_i$, 
in contrast to the case in the 2HDM where a single parameter ($\tan\beta$) determines these three couplings.
It is conventional to consider the phenomenology of the lightest $H^\pm$, assuming that the other 
$H^\pm$ are heavier. One then drops the $i$ subscript on the 
couplings of the lightest $H^\pm$ and uses the Lagrangian in eq.~(\ref{lagrangian}).

Many experimental observables in  flavour physics would receive a contribution from $H^\pm$, and thus 
the magnitudes of $X$, $Y$ and $Z$ are constrained.
Detailed studies have been performed in \cite{Grossman:1994jb}
(in the context of a MHDM) and more recently in \cite{Jung:2010ik} in the context of
the A2HDM (see also \cite{Eriksson:2008cx}).
For $m_{H^\pm} < m_t$ these constraints are roughly as follows: 
$|Y|< 1$ from $Z\to b\overline b$ (assuming $|X|<50$), $|Z|< 40$ from leptonic $\tau$ decays, and $|XZ|<1080$ from $B^\pm \to \tau\nu$.
In this work we will derive constraints on $|X|$ from $t\to H^\pm b$.

A particularly important constraint on the mass and couplings of $H^\pm$ in a 2HDM/MHDM is the decay $b\to s \gamma$ 
\cite{Hou:1987kf,Borzumati:1998tg}, which has been
measured to be in agreement with the SM prediction. It is the
combination of couplings $XY^*$ and $|Y|^2$ which enters the decay rate for $b\to s \gamma$ 
(the contribution from $|X|^2$ can be neglected).
Since $XY^*=1$ in Model II and the flipped 2HDM (i.e. $d\ne u$),
the stringent bound $m_{H^\pm}> 295$ GeV at 95\% c.l 
can be derived for all values of $\tan\beta$ 
\cite{Misiak:2006zs}. In contrast, in the MHDM with $d\ne u$, the combination $XY^*$ is only weakly constrained by 
the sum rule in eq.~(\ref{sumruleXY}), and can be negative. For $d=u$ (e.g. Model I and the leptonic-specific 2HDM) the constraint 
in eq.~(\ref{sumruleXY}) does not apply. 
Therefore a light $H^\pm$ (i.e. $m_{H^\pm} < m_t$) being compatible with $b\to s \gamma$ is still a possibility.
Recent studies of the bounds
on $H^\pm$ of the MHDM from $b\to s \gamma$ \cite{Jung:2010ik,Trott:2010iz} derive the following 
approximate $2\sigma$ intervals for the real part of $XY^*$ with $m_{H^\pm}=100$ GeV:
\begin{equation}
-1.1 < {\rm Re}\, XY^* < 0.7.
\label{bsy_xy}
\end{equation}

In deriving this constraint it is assumed that $|Y|$ is not so big (e.g. $|Y|<1$, which is
required from other low-energy processes such as $Z\to b\overline b$).
There is also a constraint on the imaginary part of $XY^*$ from a different process
(${\rm Im}\, XY^* < 0.1$), but for simplicity we will consider 
$X$ and $Y$ to be real.

In addition to the above phenomenological constraints, there are constraints on $X$, $Y$ and $Z$ from the unitarity of the matrix $U$ 
for the democratic 3HDM, which were studied in \cite{Grossman:1994jb,Cree:2011uy}. There is a non-trivial relationship on the couplings
of the lightest $H^\pm$ given as:
\begin{equation}
|X_2|^2|U_{1d}^2|+|Y_2|^2|U_{1u}^2|+|Z_2|^2|U_{1e}^2|=1
\label{XYZconstraint}
\end{equation}
This constraint ensures that the magnitudes of $X_2$, $Y_2$ and $Z_2$ cannot all be simultaneously less than one, or
all be simultaneously greater than one. This is due to the fact that all three vacuum expectation values 
$(v_d, v_u, v_e)$ cannot be simultaneously large or small.
 
In our numerical analysis we will always take $|Y|<0.8$ (as discussed above), 
and so the requirement that all three couplings cannot be simultaneously greater than unity is automatically satisfied.
We will be concerned with the parameter space of $|X|>> |Y|,|Z|$ (which corresponds to large $\tan\beta$ and
moderate/small $\tan\gamma$), a choice which satisfies the requirement
that all three couplings cannot be simultaneously less than unity.
In Fig.~(\ref{XYZconstraint}) we show the region in the plane $[X,Y,Z]$ allowed by the
unitarity constraint in eq.~(\ref{XYZconstraint}), imposing $|Y|<0.3$ and
the constraint on $XY^*$ in eq.~(\ref{bsy_xy}). It can be seen that there is no parameter space where both  
of $|X|$ and $|Z|$ are less than unity, and the parameter of interest to us (i.e. $|X|>> |Y|,|Z|$) is
fully compatible with the unitarity constraint in eq.~(\ref{XYZconstraint}).
This constraint on the couplings is removed in a 4HDM due to the presence of a fourth vacuum expectation value.
\begin{figure}[t]
\begin{center}
\includegraphics[origin=c, angle=0, scale=0.7]{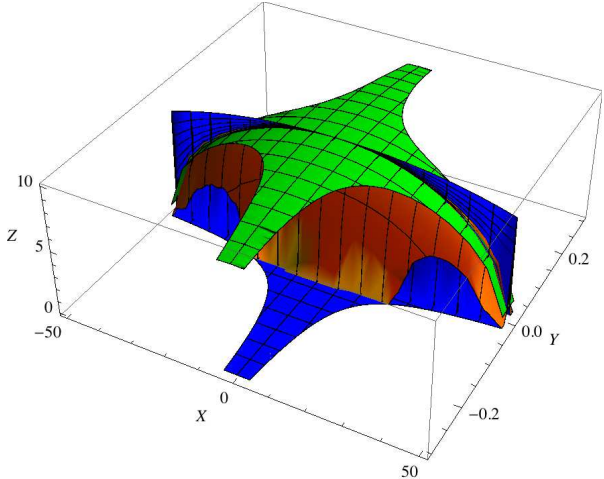}
\vspace*{-5mm}
\caption{The region of the $[X,Y,Z]$ plane allowed by the unitarity constraint
of eq.~(\ref{XYZconstraint}).
The constraint from $b\to s\gamma$ is shown as $|XY^*| < 1.1$
for Re$(XY^*) < 0$, and $|XY^*| < 0.7$ for Re$(XY^*) >0$.}
\label{XYZcontraint}
\end{center}
\end{figure}
 
\subsection{The Aligned 2HDM (A2HDM)}

The A2HDM is a 2HDM in which NFC is not imposed \cite{Pich:2009sp}, and both scalar doublets 
($\Phi_1$ and $\Phi_2$) couple
to all types of fermions. Tree-level FCNCs 
are eliminated by imposing an alignment of the Yukawa couplings of $\Phi_1$ and $\Phi_2$.
The interaction of $H^\pm$ with the fermions in 
the A2HDM can be written in the same way as eq.~(\ref{lagrangian}), but the couplings $X$, $Y$,  and $Z$
are determined by five parameters (instead of the four parameters in the democratic 3HDM) and the unitarity constraint
of eq.~(\ref{XYZconstraint}) does not apply. 
Apart from these two differences (which were discussed in \cite{Cree:2011uy}), 
the phenomenology
of $H^\pm$ in the democratic 3HDM and the A2HDM is essentially the same, and our numerical results will apply
equally to both models. In particular, the magnitudes  
of $X$,$Y$ and $Z$ determine the BRs of $H^\pm$. In the A2HDM the 
extra free parameter can be taken to be a phase in the coupling $Y$, and such a phase does not have an effect on the
BRs of $H^\pm$. Moroever, we will be concerned with the parameter space of $|X|>> |Y|,|Z|$, which is compatible
with the unitarity constraint in a 3HDM. Hereafter, when the text refers to ``$H^\pm$ of the MHDM'', the implicit
meaning is for an $H^\pm$ of a 3HDM, the A2HDM, and for a MHDM with more than three scalar doublets.

\section{A large BR($H^{\pm}\to cb)$ in the MHDM and A2HDM}

In a MHDM and in the A2HDM the expressions for the partial widths of the decay modes
of $H^\pm$ are:
\begin{equation}
\Gamma(H^\pm\to \ell^\pm\nu)=\frac{G_F m_{H^\pm} m^2_\ell |Z|^2}{4\pi\sqrt 2} 
\label{width_tau}
\end{equation}
\begin{equation}
\Gamma(H^\pm\to ud)=\frac{3G_F m_{H^\pm}(m_d^2|X|^2+m_u^2|Y|^2)}{4\pi\sqrt 2}
\label{width_ud}
\end{equation}

In $\Gamma(H^\pm\to ud)$ the running quark masses should be evaluated at the scale of $m_{H^\pm}$, and
there are QCD vertex corrections which multiply the above partial widths by
$(1+17\alpha_s/(3\pi))$. In the 2HDM the parameter $\tan\beta$ determines the magnitude of the 
partial widths. The branching ratios are well known, and for the case of interest of $m_{H^\pm}< m_t$ 
one finds that the dominant decay mode
is either $H^\pm \to cs$ or  $H^\pm \to \tau\nu$, depending on the value of $\tan\beta$.
In model I the BRs are independent of $\tan\beta$, and BR($H^\pm \to \tau\nu$)
is about twice that of BR($H^\pm \to cs$). 
%Note that the scenario of $m_{H^\pm}< m_t$ in Model II and the flipped 2HDM violates  
%the constraint from $b\to s\gamma$.

The magnitude of BR($H^\pm \to cb$) is always less than a few 
percent in three (Models I, II and lepton-specific) of the four versions of the 2HDM with NFC, since the
decay rate is suppressed by the small CKM element $V_{cb}\, (\ll V_{cs})$.
In contrast, a sizeable BR($H^\pm \to cb$) can be obtained in
the flipped 2HDM for $\tan\beta > 3$. This possibility was not stated explicitly in \cite{Barger:1989fj} when 
the flipped 2HDM was discussed. The first explicit mention of a large BR($H^\pm \to cb$) seems to have been in \cite{Grossman:1994jb},
and a quantitative study followed soon afterwards in \cite{Akeroyd:1994ga}.
As discussed in section II, the condition $m_{H^\pm}< m_t$ in the flipped 2HDM
would require additional New Physics in order to avoid the constraint on $m_{H^\pm}$
from $b\to s \gamma$, while this is not the case in the MHDM.

\subsection{The dominance of BR($H^\pm\to cb$) for  $|X|>> |Y|,|Z|$}
A distinctive signal of $H^\pm$ from a MHDM for $m_{H^\pm}< m_t$ would be a 
sizeable branching ratio for $H^\pm \to cb$. 
For $m_{H^\pm} < m_t$, the scenario of $|X|>> |Y|,|Z|$ in a MHDM 
gives rise to a ``leptophobic'' $H^\pm$ with 
BR$(H^\pm\to cs)$+BR$(H^\pm\to cb) \sim 100\%$. Consequently, BR($H^\pm \to \tau\nu$)
is negligible ($<<1\%$). The other decays of $H^\pm$ to quarks are subdominant, 
with BR($H^\pm\to us)\sim 1\%$ and BR($H^\pm\to t^*b$) only becomes
sizeable for $m_{H^\pm}\sim m_t$, as can be seen in the numerical analysis in 
\cite{Logan:2010ag} in the flipped 2HDM. Note that the 
case of $|X|>> |Y|,|Z|$ is obtained in the flipped 2HDM
for $\tan\beta > 3$, because $|X|= \tan\beta=1/|Y|=1/|Z|$ in this model.

In the scenario of $|X|>> |Y|,|Z|$
the ratio of the two dominant decays, BR$(H^\pm\to cb)$ and BR$(H^\pm\to cs)$, approaches a constant value,
which is given as follows:
\begin{equation}
\frac{{\rm BR}(H^\pm\to cb)}{{\rm BR}(H^\pm\to cs)}=R_{bs}\sim \frac{|V_{cb}|^2 m_b^2}{|V_{cs}|^2 m_s^2}
\label{cbcsratio}
\end{equation}
The CKM elements are well measured, with $V_{cb}\sim 0.04$ (a direct measurement) and $V_{cs}\sim 0.97$
(from the assumption that the CKM matrix is unitary). The running quark masses $m_s$ and $m_b$ should be evaluated at the scale
$Q=m_{H^\pm}$, and this constitutes the main uncertainty in the ratio $R_{bs}$. There is relatively little uncertainty for $m_b$,
with $m_b \,(Q=100 \,{\rm GeV})\sim 3 $ GeV.
There is more uncertainty in the value of $m_s$, although in recent years there has been much progress
in lattice calculations of $m_s$, and an average of six distinct unquenched calculations 
\cite{Durr:2010vn} gives
$m_s=93.4\pm 1.1$ MeV \cite{Laiho:2009eu} at the scale of $Q=2$ GeV. 
A more conservative average of these six calculations, $m_s=94\pm 3$ MeV, is given in \cite{Colangelo:2010et}.
 In \cite{Nakamura:2010zzi} the 
currently preferred range at $Q=2$ GeV is given as $80 \,{\rm GeV} < m_s < \, 130 $ MeV. 
Using $m_s=93$ MeV at the scale of $Q=2$ GeV (i.e. roughly the central value of the lattice averages
in \cite{Laiho:2009eu,Colangelo:2010et})
one obtains $m_s \,(Q=100 \,{\rm GeV})\sim 55 $ MeV. Taking $m_s=80$ MeV and $m_s=130$ MeV 
at $Q=2$ GeV one obtains  $m_s\sim 48$ MeV and $m_s\sim 78$ MeV respectively at $Q=100$ GeV.
   
Smaller values of 
$m_s$ will give a larger  BR$(H^\pm\to cb)$, as can be seen from eq.~(\ref{cbcsratio}).
Note that the value $m_s=55$ MeV is significantly smaller than the typical values $m_s\sim 150\to 200$ MeV which were often 
used in Higgs phenomenology in the past two decades. 
We emphasise that the scenario of  $|X|>> |Y|,|Z|$ with $m_{H^\pm}< m_t$ has a unique feature that 
the magnitude of $m_s$ is crucial for determining the relative magnitude of the two dominant decay
channels of $H^\pm$. This is not the case for most other non-minimal Higgs sectors with $H^\pm$ that are commonly studied in the literature.

In \cite{Akeroyd:1994ga} the magnitude of BR$(H^\pm\to cb)$ in the MHDM was studied in the plane of
$|X|$ and $|Y|$, for $|Z|=0$ and $0.5$, taking $m_s=0.18$ GeV and $m_b=5$ GeV.
With these quark masses the maximum value is $R_{bs}=1.23$, which corresponds to BR$(H^\pm\to cb)\sim 55\%$.
However, the values of $m_s=0.18$ GeV and $m_b=5$ GeV  are not realistic 
(as was subsequently noted in \cite{Akeroyd:1998dt}), and 
two recent papers \cite{Logan:2010ag,Aoki:2009ha} have updated the magnitude of $R_{bs}$ in the flipped 2HDM
using realistic running quark masses at the scale of $m_{H^\pm}$.
In \cite{Logan:2010ag}, it appears that $m_s=0.080$ GeV at the scale of $m_{H^\pm}$ 
was used, which gives BR$(H^\pm\to cb)\sim 70\%$, in agreement with our results.
In \cite{Aoki:2009ha}, $m_s=0.077$ GeV at the scale of $m_{H^\pm}$ 
was used, with a maximum value for BR$(H^\pm\to cb)$ of $\sim 70\%$. 
We note that none of these papers used the precise average $m_s=93.4\pm 1.1$ MeV \cite{Laiho:2009eu}
of the lattice calculations, which gives $m_s\sim 55$ MeV at the scale of
 $m_{H^\pm}$. This smaller value of $m_s$ leads to a maximum value of 
BR$(H^\pm\to cb)$ which is larger than
that given in \cite{Akeroyd:1994ga,Aoki:2009ha,Logan:2010ag}, as discussed below.

\begin{figure}[t]
\begin{center}
\includegraphics[origin=c, angle=0, scale=0.5]{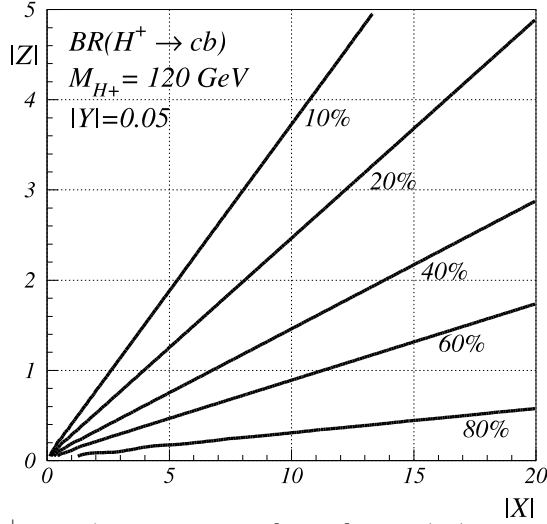}
\vspace*{-5mm}
\caption{Contours of BR$(H^\pm\to cb$) in the plane [$X$, $Z$] with $|Y|=0.05$.
We take $m_s(Q=m_{H^\pm})=0.055$ GeV and $m_{H^\pm}=120$ GeV.}
\label{fig:brcb_xz}
\end{center}
\end{figure}

We now study the magnitude of $H^\pm \to cb$ 
as a function of the couplings $X,Y,Z$.
In Fig.~(\ref{fig:brcb}a) we update the numerical study of \cite{Akeroyd:1994ga} for
BR$(H^\pm\to cb)$ in the plane $[X,Y]$ in a MHDM with $|Z|=0.1$, using $m_s=0.055$ GeV and $m_b=2.95$ GeV
at the scale of $m_{H^\pm}=120$ GeV.
With these values for the quark masses the maximum value is BR$(H^\pm\to cb)\sim 81\%$ 
i.e. a significantly larger value than BR$(H^\pm\to cb)\sim 55\%$ in \cite{Akeroyd:1994ga}. 
Taking a lower value of $m_s=0.08$ GeV one has BR$(H^\pm\to cb)\sim 69\%$, and for 
$m_s=0.048$ GeV one has BR$(H^\pm\to cb)\sim 86\%$.
In Fig.~(\ref{fig:brcb}a) we also display the bound from $b\to s \gamma$ (for $m_{H^\pm}=100$ GeV), 
which is $|XY| < 1.1$ for $XY^*$ being real and negative, and
$|XY| < 0.7$ for $XY^*$ being real and positive. The parameter space for 
BR$(H^\pm\to cb)> 60\%$ roughly corresponds to $|X| > 1$ and $|Y| < 0.25$ for $|XY| < 0.7$.
In  Fig.~(\ref{fig:brcb}b) and   Fig.~(\ref{fig:brcb}c) we show BR$(H^\pm\to cs)$ 
and  BR$(H^\pm\to \tau\nu)$ respectively. As expected, BR$(H^\pm\to cs)$ is maximised for
$|Y|>> |X|,|Z|$ while  BR$(H^\pm\to \tau\nu)$ is maximised for  $|Z|>> |X|,|Y|$.   
In Fig.~(\ref{fig:brcb_xz}) we show contours of BR$(H^\pm\to cb)$ in the plane $[X,Z]$ for $m_{H^\pm}=120$ GeV
and $|Y|=0.05$. For this value of $|Y|$ the constraint from $b\to s\gamma$ is always satisfied
for the displayed range of $|X|<20$. One can see that the largest values of BR$(H^\pm\to cb)$
arise for $|Z|< 2$.

\subsection{The decay $H^\pm \to A^0W^*$ for $m_{A^0}< m_{H^\pm}$}
The above discussion has assumed that $H^\pm$ cannot decay into other scalars.
We now briefly discuss the impact of the decay channel $H^\pm\to A^0W^*$, which has been studied in
the 2HDM (Type II) in \cite{Moretti:1994ds} and in other 2HDMs with small $|X|,|Y|$ and $|Z|$ 
in \cite{Akeroyd:1998dt}, with direct searches at LEP (assuming $A^0\to b\overline b$) performed in \cite{Abdallah:2003wd}.
In a general non-SUSY 2HDM the masses of the scalars can be taken as free parameters. This is in contrast
to the MSSM in which one expects $m_{H^\pm}\sim m_{A^0}$ in most of the parameter space.
The scenarios of $m_{A^0}< m_{H^\pm}$ and $m_{A^0}> m_{H^\pm}$ are both possible in a 2HDM, but 
large mass splittings among the scalars lead to sizeable contributions to electroweak precision observables
\cite{Toussaint:1978zm}, which
are parametrised by the $S$, $T$ and $U$ parameters \cite{Peskin:1990zt}.
The case of exact degeneracy ($m_{A^0}=m_{H^0}=m_{H^\pm}$)
leads to values of $S$, $T$ and $U$ which are almost identical to those of the SM.
A recent analysis in a 2HDM \cite{Kanemura:2011sj} 
sets $m_{H^0}=m_{A^0}$ and $\sin(\beta-\alpha$)=1, and studies the maximum value of the 
mass splitting $\Delta m= m_{A^0}-m_{H^\pm}$ (for earlier studies see \cite{Chankowski:1999ta}). 
For $m_{A^0}=100$ GeV
the range $-70 \,{\rm GeV} < \Delta m < 20 \,{\rm GeV}$ is 
allowed, which corresponds to $80 \,{\rm GeV} < m_{H^\pm} < 170 \,{\rm GeV}$. 
For $m_{A^0}=150$ GeV the allowed range is
$-70 \,{\rm GeV} < \Delta m < 70 \,{\rm GeV}$ which corresponds to $80 \,{\rm GeV} < m_{H^\pm} < 220 
\,{\rm GeV}$. Consequently, sizeable mass splittings (of either sign) of the scalars are possible.
Analogous studies in a MHDM have been performed in \cite{Grimus:2008nb}, with similar conclusions.

\begin{figure}[t]
\begin{center}
\includegraphics[origin=c, angle=0, scale=0.5]{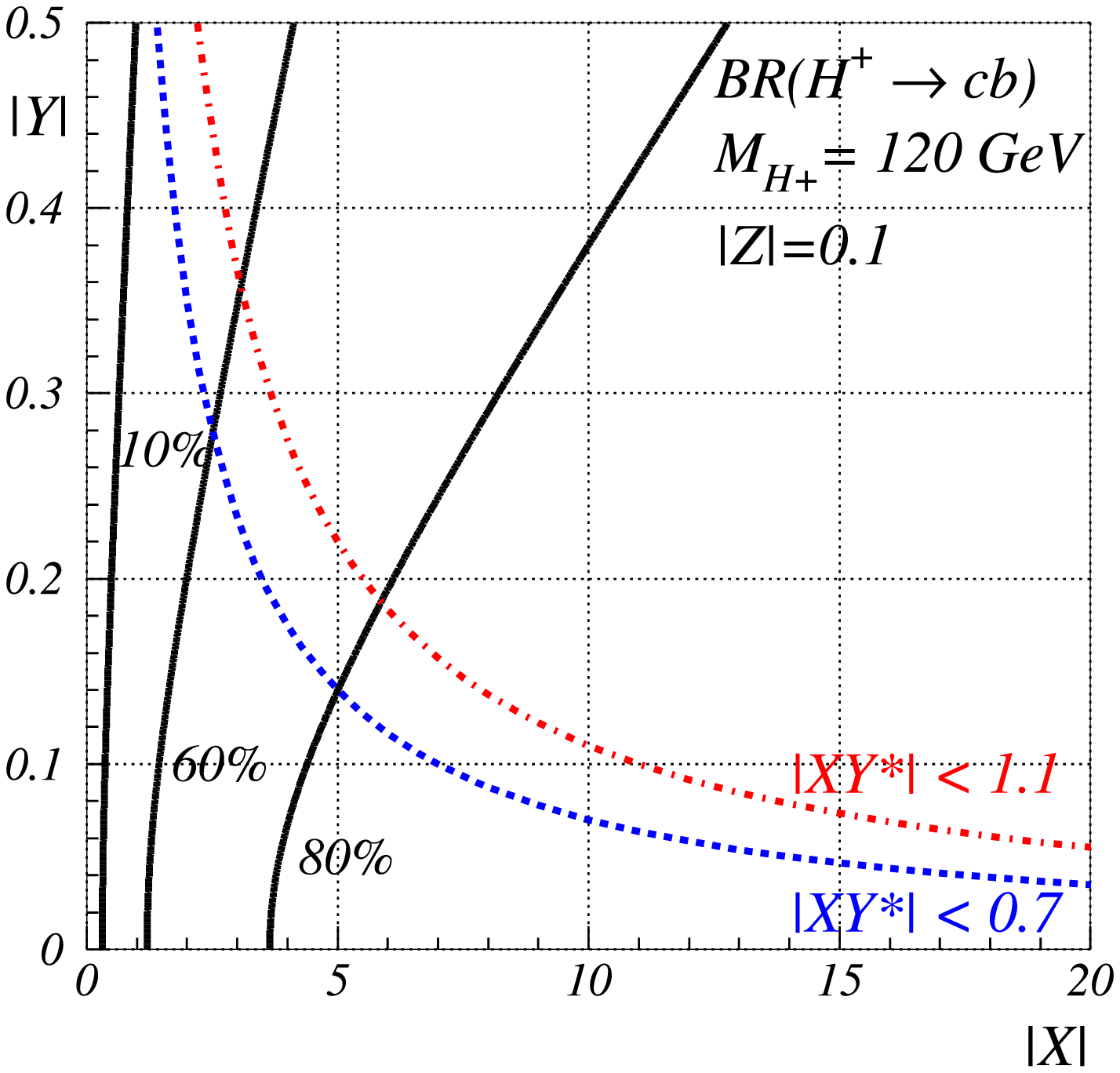}
\includegraphics[origin=c, angle=0, scale=0.5]{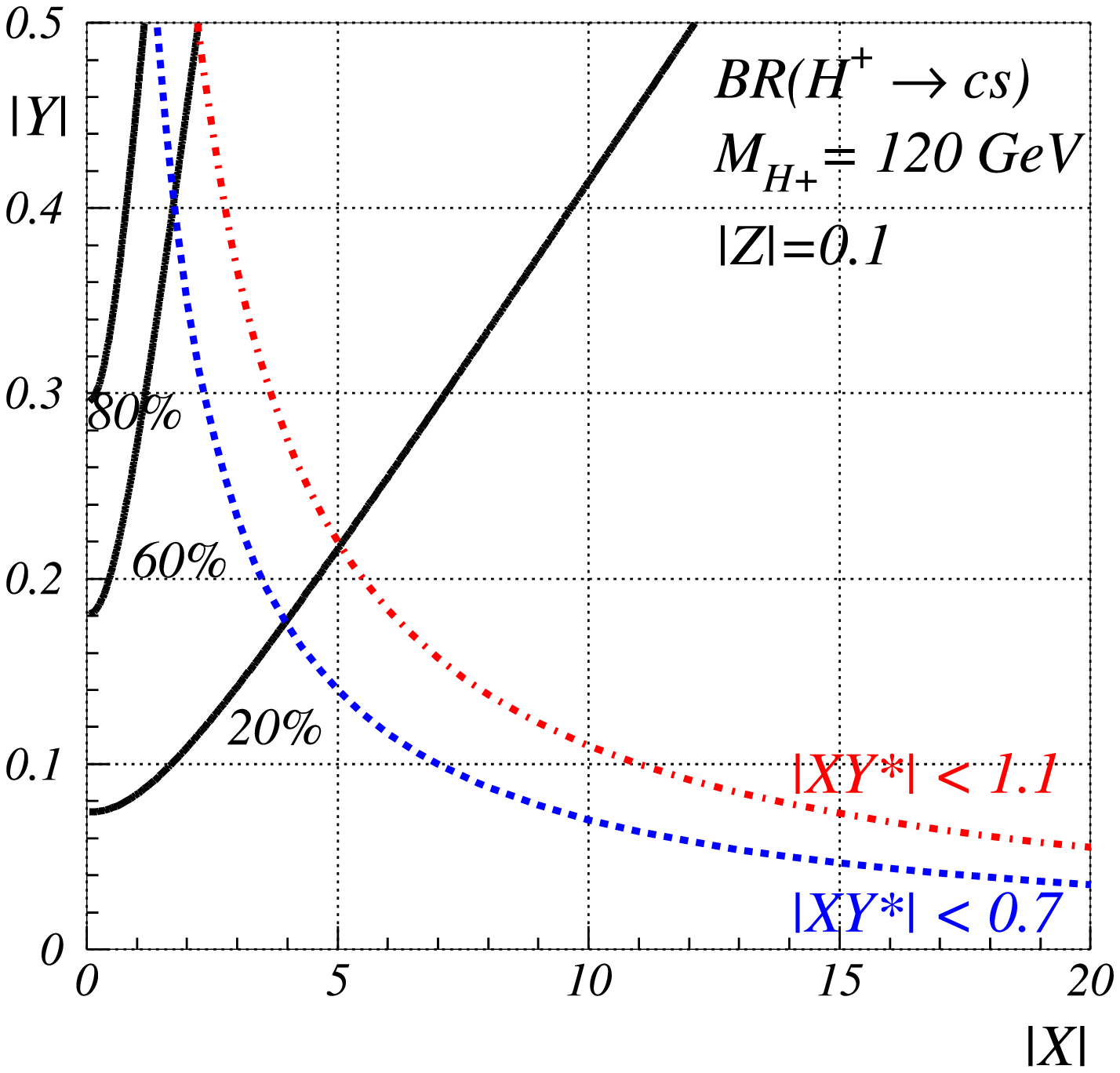}
\includegraphics[origin=c, angle=0, scale=0.5]{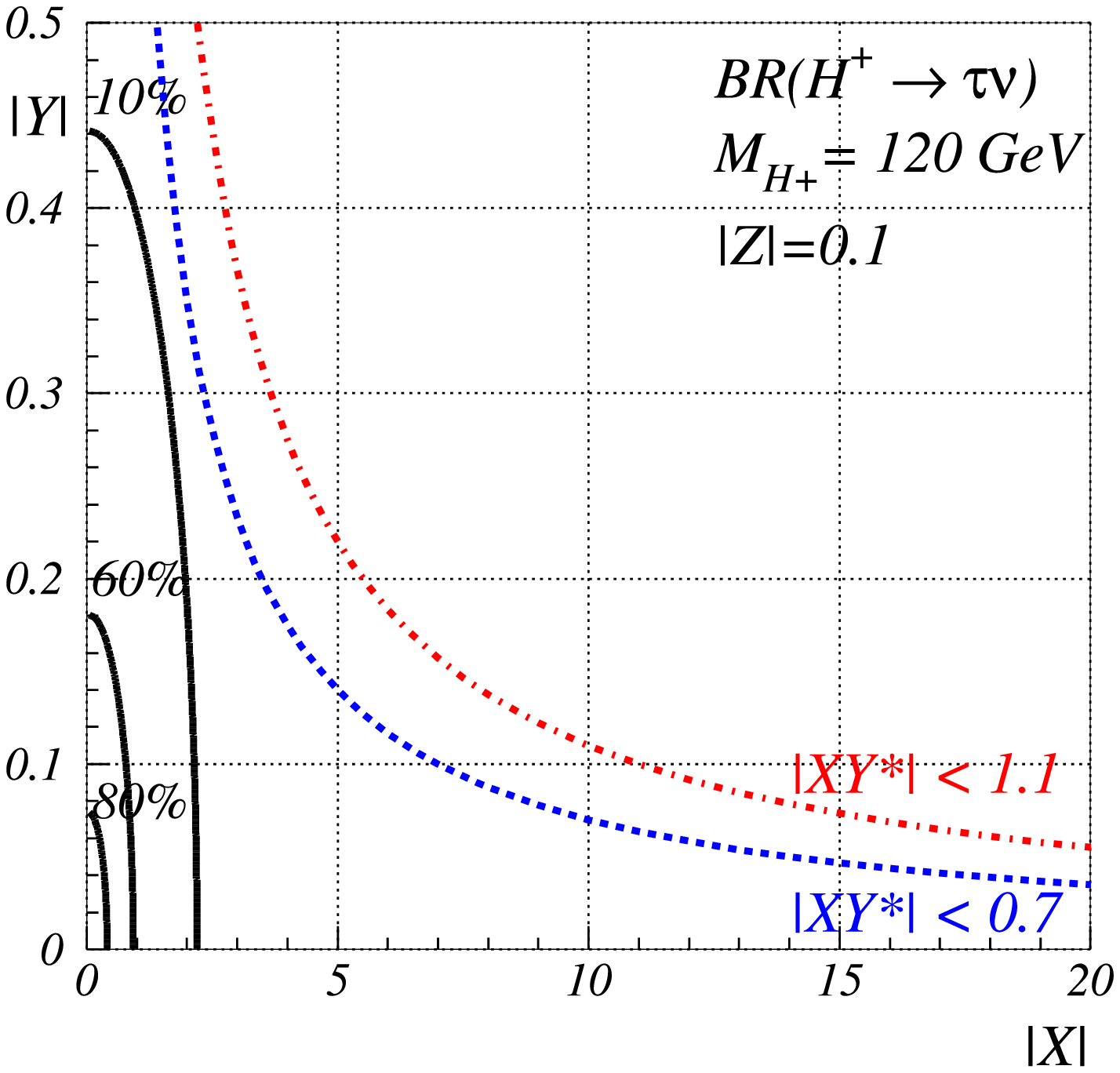}
\vspace*{-5mm}
\caption{Contours of BR$(H^\pm\to cb$), BR$(H^\pm\to cs$) and 
 BR$(H^\pm\to \tau\nu$) in the plane [$X$, $Y$] with $|Z|=0.1$.
The constraint from $b\to s\gamma$ is shown as $|XY^*| < 1.1$
for Re$(XY^*) < 0$, and $|XY^*| < 0.7$ for Re$(XY^*) >0$.
We take $m_s(Q=m_{H^\pm})=0.055$ GeV and $m_{H^\pm}=120$ GeV. }
\label{fig:brcb}
\end{center}
\end{figure}

\begin{figure}[t]
\begin{center}
\includegraphics[origin=c, angle=0, scale=0.5]{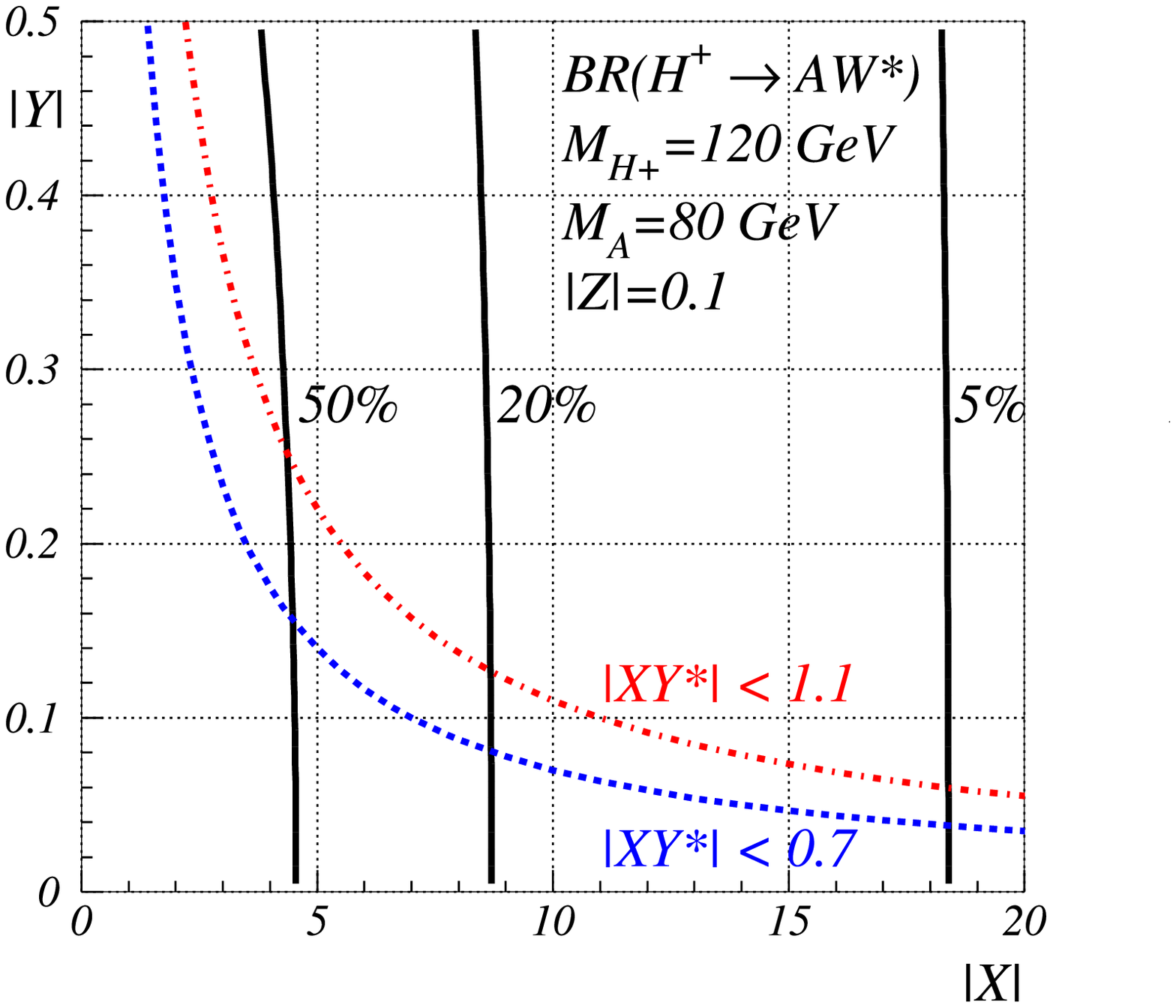}
\includegraphics[origin=c, angle=0, scale=0.5]{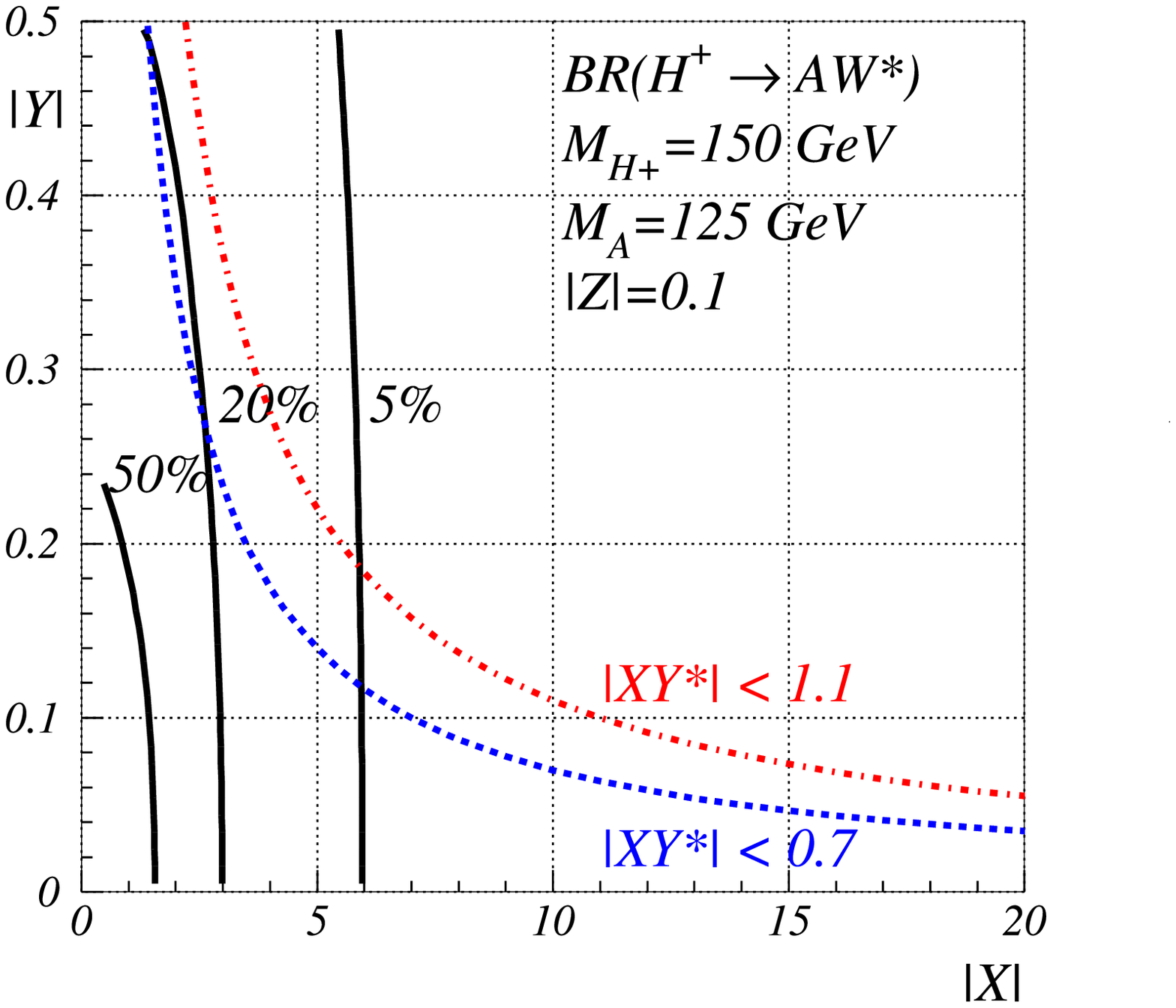}
\vspace*{-5mm}
\caption{Contours of BR$(H^\pm\to A^0W^*$) in the plane [$X$, $Y$] with $|Z|=0.1$.
The constraint from $b\to s\gamma$ is shown as $|XY^*| < 1.1$
for Re$(XY^*) < 0$, and $|XY^*| < 0.7$ for Re$(XY^*) >0$. 
In the left panel (a) we take $m_{H^\pm}=120$ GeV and $m_{A^0}=80$ GeV, and in the right panel (b)
we take $m_{H^\pm}=150$ GeV and $m_{A^0}=125$ GeV. In both figures
$m_s(Q=m_{H^\pm})=0.055$ GeV.}
\label{fig:braw}
\end{center}
\end{figure}

If $m_{A^0}< m_{H^\pm}$ then the decay channel $H^\pm\to A^0W^*$ can compete with the
above decays of $H^\pm$ to fermions, because the coupling $H^\pm A^0 W$ is
not suppressed by any small parameter. In Fig.~(\ref{fig:braw}a) we show contours of BR($H^\pm\to A^0W^*$)
in the plane $[X,Y]$ with $|Z|=0.1$, $m_{A^0}=80$ GeV and $m_{H^\pm}=120$ GeV.
The contours are essentially vertical
in the parameter space of interest (i.e. $|Y|<0.5$ and $|X|>> 1$) because the contribution of the 
term $m^2_c |Y|^2$ to the decay widths of $H^\pm$ to fermions is small. Comparing Fig.~(\ref{fig:braw}a) and  
Fig.~(\ref{fig:brcb}a) one can see
that for $|X|\sim 5$ both BR$(H^\pm\to A^0W^*$) and BR$(H^\pm\to cb)$ are dominant, with roughly equal BRs.
For smaller $m_{A^0}$ (e.g. $< 80 \,{\rm GeV}$) the contour of BR$(H^\pm\to A^0W^*$)=$50\%$
would move to higher values of $|X|$. 
Since the dominant decay of $A^0$ is expected to be $A^0\to b\overline b$, the detection prospects in this channel
should also be promising because there would be more $b$ quarks from
$t\to H^\pm b$, $H^\pm \to A^0W^*$, $A^0\to b\overline b$ than from $t\to H^\pm b$ with $H^\pm \to cb$. 
We note that there has been a search by the Tevatron for the channel $t\to H^\pm b$, $H^\pm \to A^0W^*$, $A^0\to \tau^+\tau^-$
\cite{Tev_search_HAW}, for the case
of $m_{A^0}<2m_b$ where $A^0\to b\overline b$ is not possible \cite{Dermisek:2008uu}.

At present there is much speculation about an excess of events around a mass of 125 GeV in the
search for the SM Higgs boson \cite{latest_LHC}. An interpretation of these events as 
originating from the process
$gg\to A^0 \to \gamma\gamma$ has been suggested in \cite{Burdman:2011ki}. In Fig.~(\ref{fig:braw}b)
we set $m_{A^0}=125$ GeV and $m_{H^\pm}=150$ GeV. Since the mass splitting between $H^\pm$ and $A^0$ is less than in
Fig.~(\ref{fig:braw}a), the contours move to lower values of $|X|$, but BR$(H^\pm\to A^0W^*$)=$50\%$ is still possible
for $|X|<2$.
We note that if the excess of events at 125 GeV is attributed to a SM-like Higgs, then in the context of a 2HDM
a candidate would be the lightest CP-even Higgs $h^0$ with a coupling to vector bosons of SM strength 
(recent studies of this possibility can be found in 
\cite{Ferreira:2011aa}). This scenario would correspond to $\sin(\beta-\alpha)\sim 1$ in a 2HDM, 
and therefore the coupling $H^\pm h^0 W$ (with a magnitude $\sim \cos(\beta-\alpha$) in a 2HDM) would be close
to zero. Hence the decay $H^\pm \to h^0 W^*$ would be suppressed by this small 
coupling, as well as by the virtuality of $W^*$. Several recent studies 
\cite{Carmi:2012yp, Gabrielli:2012yz} fit the current data in all the Higgs search channels
to the case of a neutral Higgs boson with arbitrary couplings. A SM-like Higgs boson gives a good fit to the
data, although a slight preference for non-SM like couplings is emphasised in \cite{Gabrielli:2012yz}. 
If the excess of events at 125 GeV turns out to be genuine {\it and} is well described by a non-SM like Higgs boson
of a 2HDM with a value of $\sin(\beta-\alpha)$ which is significantly less than unity, then
BR($H^\pm \to h^0 W^*$) could be sizeable, with a magnitude given by Fig.~(\ref{fig:braw}b) 
after scaling by $\cos^2(\beta-\alpha)$.

\section{Searches for $t\to H^\pm b$ with $H^\pm \to cs$, and prospects for $H^\pm \to cb$ at the LHC }

The case of $m_{H^\pm} < m_t + m_b$ with a large BR($H^\pm \to cs$) 
 can be searched for in the decays of the
top quark via $t\to H^\pm b$ \cite{Barger:1989fj,Barnett:1992ut}. 
The first discussion of $t\to H^\pm b$ followed by the decay 
$H^\pm \to cb$ was given in \cite{Akeroyd:1995cf}. Recently,  $t\to H^\pm b$
with decay  $H^\pm \to cb$ has been
studied in the context of flipped 2HDM \cite{Logan:2010ag}, and in the context of the 2HDM without natural flavor conservation 
\cite{DiazCruz:2009ek}. 

There have been two dedicated searches by the Tevatron collaborations
\cite{:2009zh, Aaltonen:2009ke} for  $t\to H^\pm b$ followed by
 $H^\pm \to cs$. 
The D0 analysis \cite{:2009zh} with 1 fb$^{-1}$ performed a search 
for $t\to H^\pm b$ by studying 
the effect of the decay $H^\pm \to cs$ on ratios of cross sections for $t\overline t$
production. In the SM one has BR($t\to Wb$)= 100\%,  and the branching ratios
of $W\to \ell\nu$ and  $W\to q'\overline q$ are known. The
presence of a sizeable BR($t\to H^\pm b$) with $H^\pm \to cs$  would change the 
SM prediction for the ratio of the cross sections for the channels with 
decay $W\to \ell\nu$ and  $W\to q'\overline q$.
For the optimum case of BR($H^\pm \to cs)=100\%$, upper bounds on
BR$(t\to H^\pm b)$ between 0.19 and 0.22 were obtained for
$80 \,{\rm GeV}< m_{H^\pm} < 155$ GeV. 
Although the decay $H^\pm \to cs$ was assumed in \cite{:2009zh} 
the above limits also apply (to a very good approximation) to the case
of both $H^\pm \to cs$ and
$H^\pm \to cb$ having sizeable BRs, as discussed in \cite{Logan:2010ag}.
 This is because the search strategy merely requires that 
$H^\pm$ decays to quark jets. 

An alternative strategy was adopted in the CDF analysis \cite{Aaltonen:2009ke}
with 2.2 fb$^{-1}$. A direct search for the decay $H^\pm \to cs$ 
was performed by looking for a peak
centered at $M_{H^\pm}$ in the dijet invariant mass distribution, which would be distinct from the peak at $M_W$
from the SM decay $t\to Wb$ with $W\to q'\overline q$.  
For the optimum case of BR($H^\pm \to q'\overline q)=100\%$, upper bounds on
BR$(t\to H^\pm b)$ between 0.32 and 0.08 were obtained for
$90\, {\rm GeV}< m_{H^\pm} < 150$ GeV, with the greatest sensitivity being at $m_{H^\pm}=130$ GeV.
No limits on BR$(t\to H^\pm b)$ were given for the region $70\, {\rm GeV}< m_{H^\pm} < 90$ GeV due to the
large background from $W\to q'\overline q$ decays. For the region $60\, {\rm GeV}< m_{H^\pm} < 70$ GeV,
limits on BR$(t\to H^\pm b)$ between 0.09 and 0.12 were derived.
As stated in \cite{Aaltonen:2009ke}, the above limits also apply to
other hadronic decays of $H^\pm$, although with slight changes in the sensitivity
to BR$(t\to H^\pm b)$ because the dijet mass resolution depends mildly on the flavour of the quarks.
The search strategy in \cite{Aaltonen:2009ke} and does not have sensitivity to the
region  $80\, {\rm GeV} < m_{H^\pm} < 90$ GeV due to the large background from $W\to cs,ud$.
The combination of the four searches at LEP for $e^+e^- \to H^+H^-$ \cite{Searches:2001ac} derived the limit $m_{H^\pm}>81$ GeV for the
 scenario of BR($H^\pm\to cs)\sim 100\%$,
with the following additional small intervals excluded (at $95\%$ c.l): $86 \,{\rm GeV} < m_{H^\pm} < 88$ GeV and  $m_{H^\pm} \sim 84$ GeV. 
Therefore the region of $80\, {\rm GeV} < m_{H^\pm} < 90$ GeV with $H^\pm$ decaying
dominantly to quarks (e.g. $cs,cb$) has not yet been entirely excluded yet.

Concerning the prospects at the LHC,
there has been a simulation of $t\to H^\pm b$ followed by the decay $H^\pm \to cs$
by the ATLAS collaboration in \cite{Ferrari:2010zz}, assuming that
one of the top/antitop quarks in the $t\overline t$ events decays leptonically via $t\to Wb\to \ell\nu b$.
This strategy is very similar to the CDF analysis of
\cite{Aaltonen:2009ke}, and directly looks for a peak centered at $m_{H^\pm}$ 
in the invariant mass distribution of 
the jets from $H^\pm\to cs$. Two $b$-tags are applied, and the
peak from $H^\pm\to cs$ is obtained by reconstructing the two untagged jets.
The mass resolution of the peak  can be further
improved by full reconstruction of the $t\overline t$ event.
For $\sqrt s=7$ TeV with 1 fb$^{-1}$ of luminosity,  
values of BR$(t\to H^\pm b)$ as low as 0.04 can be probed for  $110 \,{\rm GeV}\, < m_{H^\pm} < 150 \,{\rm GeV}$.
This sensitivity is superior to that achieved for the  decay $t\to H^\pm b$
followed by $H^\pm\to \tau\nu$ with the same integrated luminosity \cite{Aaltonen:2009ke}.
Again, as in \cite{Aaltonen:2009ke} there is little or no sensitivity to the region $80\, {\rm GeV}< m_{H^\pm} < 90$ GeV.

The first search for  $t\to H^\pm b$ followed by the decay $H^\pm \to cs$ at the LHC has been performed
by the ATLAS collaboration with 0.035 fb$^{-1}$ in \cite{ATLAS:search}.
Due to the small amount of integrated luminosity, only one $b$-tag was applied.
The limits on BR$(t\to H^\pm b)$ are comparable to those from the Tevatron search in \cite{Aaltonen:2009ke}, with limits of
BR($t\to H^\pm b) <0.25, 0.15$ and $0.14$  for $m_{H^\pm}=90 \, {\rm GeV}, 110 \,{\rm GeV}$ and $130$ GeV respectively. 

If BR$(H^\pm\to cb)$ were the dominant decay channel, as can be the case in the
MHDM and the A2HDM, the requirement of tagging the $b$ from $H^\pm \to cb$ 
 (as suggested in \cite{Akeroyd:1995cf,DiazCruz:2009ek,Logan:2010ag})
would provide sensitivity to BR$(t\to H^\pm b)$
in the problematic region $80\, {\rm GeV}< m_{H^\pm} < 90$ GeV, and should improve
the sensitivity for $m_{H^\pm} > 90$ GeV. We now estimate the gain in sensitivity using realistic values
for the $b$-tagging efficiency 
($\epsilon_b=0.5$), the probability of a $c$-quark being misidentified as a $b$ quark 
($\epsilon_c=0.1$) and the 
probability of a light quark being misidentified as a $b$-quark ($\epsilon_j=0.01$).
The two dominant backgrounds to the peak at $m_{H^\pm}$ in the dijet invariant mass
distribution are from $W\to ud$ and $W\to cs$, which we take to be equal in magnitude.
For the case of BR$(H^\pm \to cb)$ near $80\%$,
the ratio of the signal to the background ($S/\sqrt B$) with and without the $b$-tag is given
approximately as follows:
\begin{equation}
\frac{[S/\sqrt B]_{\rm btag}}{[S/\sqrt B]_{\rm \not{btag}}}
\sim \frac{\epsilon_b\sqrt 2}{\sqrt{(\epsilon_j+\epsilon_c)}}\sim 2.13 \, .
\label{gain_from_btag}
\end{equation}
We encourage a detailed simulation by the Tevatron and LHC collaborations in order to 
obtain a more realistic estimate of the increase in sensitivity over the current strategy of
not applying a $b$-tag to the jets originating from $H^\pm$.

We note that a recent paper \cite{Kao:2011aa} has performed a simulation for a very similar signature which arises
from the decay $t\to h^0c\to b\overline b c$ in a different 2HDM with FCNCs. 
This signature looks identical to the signature arising from $H^\pm\to cb$ but there are several kinematical differences. 
The process $t\to H^\pm b \to b\overline b c$ would give a peak at $m_{H^\pm}$ in the dijet invariant mass distribution
in which only one of jets has originated from a $b$ quark, with the other two $b-$jets coming from the decay of $t\overline t$.  
In contrast, for $t\to h^0c\to b\overline b c$ both 
of jets in the dijet invariant mass distribution would originate from $b$ quarks,
 while the third $b-$jet would come from the decay of $t$ 
or $\overline t$.
The study in \cite{Kao:2011aa} is specifically for $t\to h^0c\to b\overline b c$, and it was found that the sensitivity to
BR($t\to h^0c$) was significantly superior to that for $t\to H^\pm b$ followed by $H^\pm\to cs$, which can be attributed to the
extra $b$-tag i.e. the increase in sensitivity in \cite{Kao:2011aa} compared to that obtained for
the LHC simulation without the $b$-tag in \cite{Ferrari:2010zz} is significantly greater than the value of 2.13 in
eq.~(\ref{gain_from_btag}), and could be as large as a factor of six.

\section{Numerical Results}

We now quantify the magnitude of 
$H^\pm \to cb$ events produced in the decays of $t$ quarks, and compare this 
with the expected sensitivity at the LHC.
For the partial decay widths of $t\to W^\pm b$ and 
$t\to H^\pm b$ we use the leading-order expressions (with $|V_{tb}|=1$) as follows: 
\begin{eqnarray}
\Gamma(t\to W^\pm b)=\frac{G_F m_t}{8\sqrt 2 \pi}[m_t^2+2M_W^2][1-M_W^2/m_t^2]^2  \\ \nonumber
\Gamma(t\to H^\pm b)=\frac{G_F m_t}{8\sqrt 2 \pi}[m^2_t|Y|^2 + m_b^2|X|^2][1-m_{H^\pm}^2/m_t^2]^2
\end{eqnarray}

The multiplicative (vertex) QCD corrections to both $t\to W^\pm b$ and 
$t\to H^\pm b$ essentially cancel out in the ratio of partial widths \cite{Li:1990cp}, and thus they
do not affect  BR$(t\to H^\pm b)$ significantly. In the phase-space function of both decays we
neglect $m_b$, and in the terms $m^2_t|Y|^2$ and $m_b^2|X|^2$
we use $m_t=175$ GeV and $m_b$ evaluated at the scale of $m_{H^{\pm}}$ (i.e. $m_b\sim 2.95$ GeV).

\begin{figure}[t]
\begin{center}
\includegraphics[origin=c, angle=0, scale=0.5]{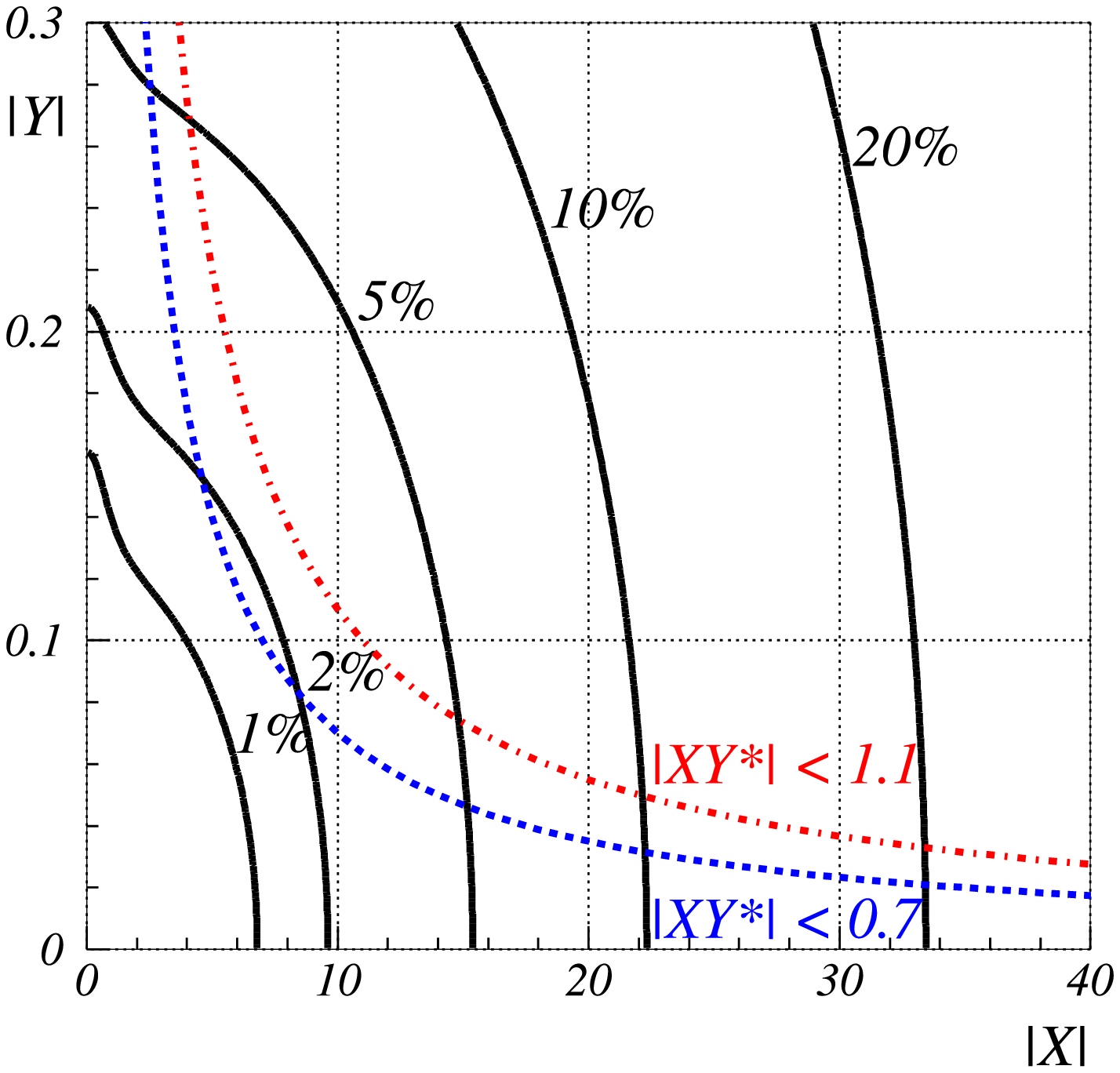}
\includegraphics[origin=c, angle=0, scale=0.5]{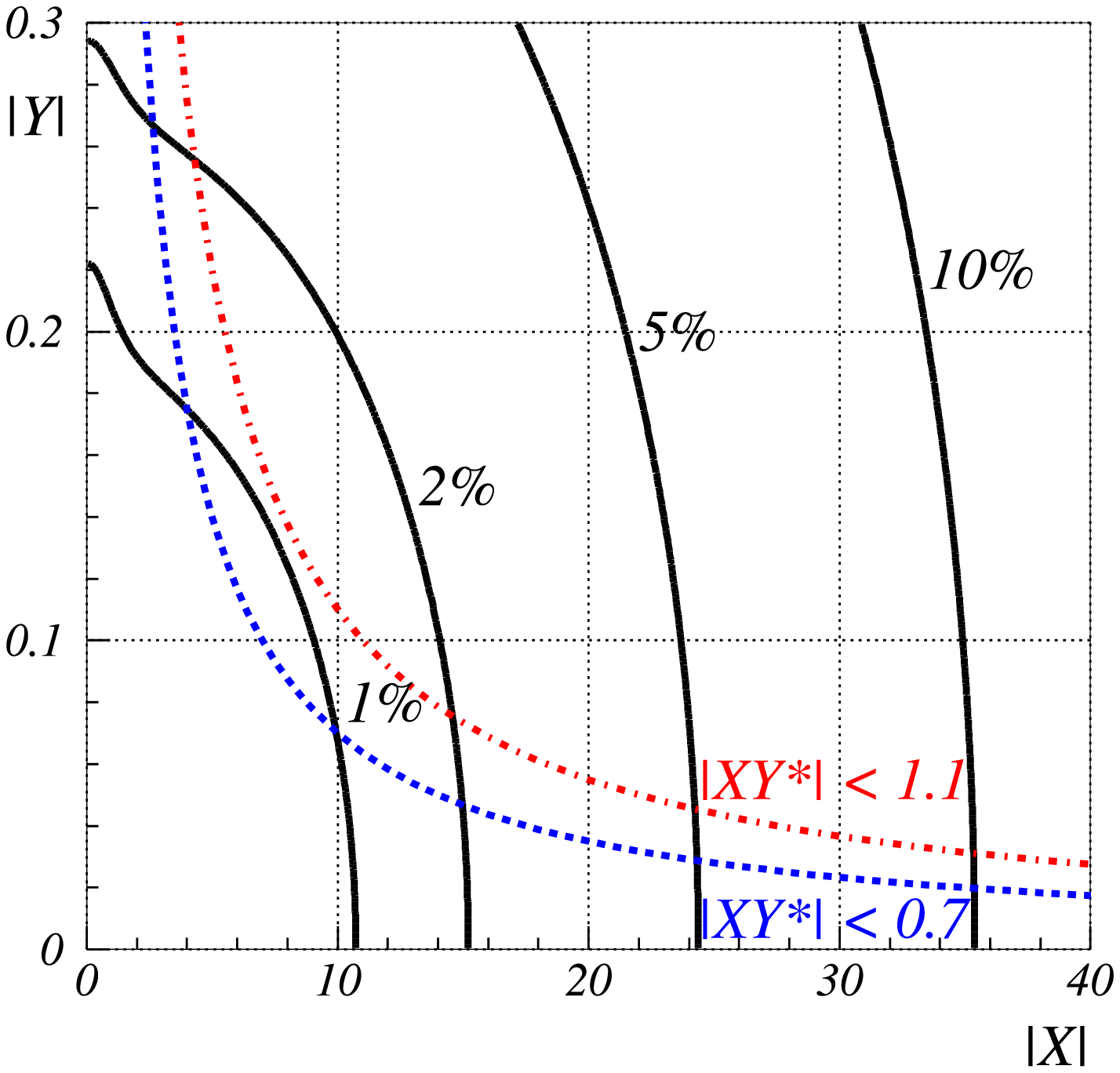}
%\vspace*{-5mm}
\caption{Contours of the sum of BR$(t\to H^\pm b)\times {\rm BR}(H^\pm\to cs$) 
and BR$(t\to H^\pm b)\times {\rm BR}(H^\pm\to cb$)
in the plane [$X$, $Y$] with $|Z|=0.1$, where $m_{H^\pm}=80$ GeV (left panel) and $m_{H^\pm}=120$ GeV (right panel).}
\label{fig:brcbcs}
\end{center}
\end{figure}

In Fig.~(\ref{fig:brcbcs}a) and Fig.~(\ref{fig:brcbcs}b) we show contours of the sum of 
\begin{equation}
{\rm BR}(t\to H^\pm b)\times [{\rm BR}(H^\pm\to cs) + {\rm BR}(H^\pm\to cb)] 
\label{cbcs}
\end{equation}
in the plane of $[X,Y]$ for $m_{H^\pm}=$80 GeV 
and $m_{H^\pm}=$120 GeV respectively, setting $|Z|=0.1$. The cross section in eq.~(\ref{cbcs})
is the signature to which the current search strategy at
the Tevatron and the LHC is
sensitive, i.e. one $b$-tag (LHC \cite{ATLAS:search}) or two $b$-tags (Tevatron \cite{Aaltonen:2009ke}) 
are applied to the jets originating from $t\overline t$ decay, but no $b$-tag is applied to the jets originating from $H^\pm$. 
For the case of [BR($H^\pm \to cs)$+BR($H^\pm \to cb)$]=100\% the current experimental limits for $m_{H^\pm}=120$ GeV are ${\rm BR}(t\to H^\pm b) <0.14$ from ATLAS
with 0.035 fb$^{-1}$ \cite{ATLAS:search},
${\rm BR}(t\to H^\pm b) <0.12$ from CDF with 2.2 fb$^{-1}$ \cite{Aaltonen:2009ke}, and  ${\rm BR}(t\to H^\pm b) <0.22$ from
D0 with 1 fb$^{-1}$ \cite{:2009zh}. In Fig.~(\ref{fig:brcbcs}b) for $m_{H^\pm}=$120 GeV these upper limits would exclude
the parameter space of $|X|>40$ and small $|Y|$ which is not excluded by the constraint from $b\to s\gamma$.
For the mass region $80\, {\rm GeV} < m_{H^\pm} < 90$ GeV there is only a limit from the D0 search in \cite{:2009zh}, which
gives BR$(t\to H^\pm b)<0.21$. From Fig.~(\ref{fig:brcbcs}a), for $m_{H^\pm}=80$ GeV 
one can see that this limit excludes the parameter space of $|X|>35$ and small $|Y|$. 

In both Fig.~(\ref{fig:brcbcs}a) and Fig.~(\ref{fig:brcbcs}b) we show contours of $1\%$, 
which might be reachable in the 8 TeV run of the LHC. 
Simulations by ATLAS (with $\sqrt s=7$ TeV) for $H^\pm\to cs$ \cite{Ferrari:2010zz} have shown that the LHC 
should be able to probe values BR$(t\to H^\pm b)>0.05$ with 1 fb$^{-1}$
for $m_{H^\pm} > 110$ GeV, with the
greatest sensitivity being around $m_{H^\pm}=130$ GeV. 
For the operation with $\sqrt s=8$ TeV and an anticipated integrated luminosity of 15 fb$^{-1}$ 
one expects increased sensitivity (e.g. BR$(t\to H^\pm b)>0.01$ for $m_{H^\pm} > 110$ GeV), 
although the region $80 \,{\rm GeV} < m_{H^\pm} < 90$ GeV might remain difficult to probe with the strategy of
reconstructing the jets from $H^\pm$. An alternative way to probe  the 
region $80 \,{\rm GeV} < m_{H^\pm} < 90$ GeV is to use
the search strategy by D0 in \cite{:2009zh}, and presumably the LHC could improve on the Tevatron 
limit on BR$(t\to H^\pm b) < 0.21$ for $80 \,{\rm GeV} < m_{H^\pm} < 90$ GeV.
From  Fig.~(\ref{fig:brcbcs}b) (for $m_{H^\pm}=120$ GeV) one can see that the region 
of $|Y|> 0.2$ and $|X|< 4$, which is not excluded by $b\to s \gamma$, would be probed
if sensitivity to BR$(t\to H^\pm b)>0.01$ were achieved.
However, a large part of the 
region roughly corresponding to $|Y|<0.2$ and $|X|< 10$ (which is also not excluded by $b\to s \gamma$) would require
sensitivity to BR$(t\to H^\pm b)< 0.01$ in order to be probed  
with the current search strategy for $t\to H^\pm b$, and this is probably unlikely in the 8 TeV run of the LHC.

\begin{figure}[t]
\begin{center}
\includegraphics[origin=c, angle=0, scale=0.5]{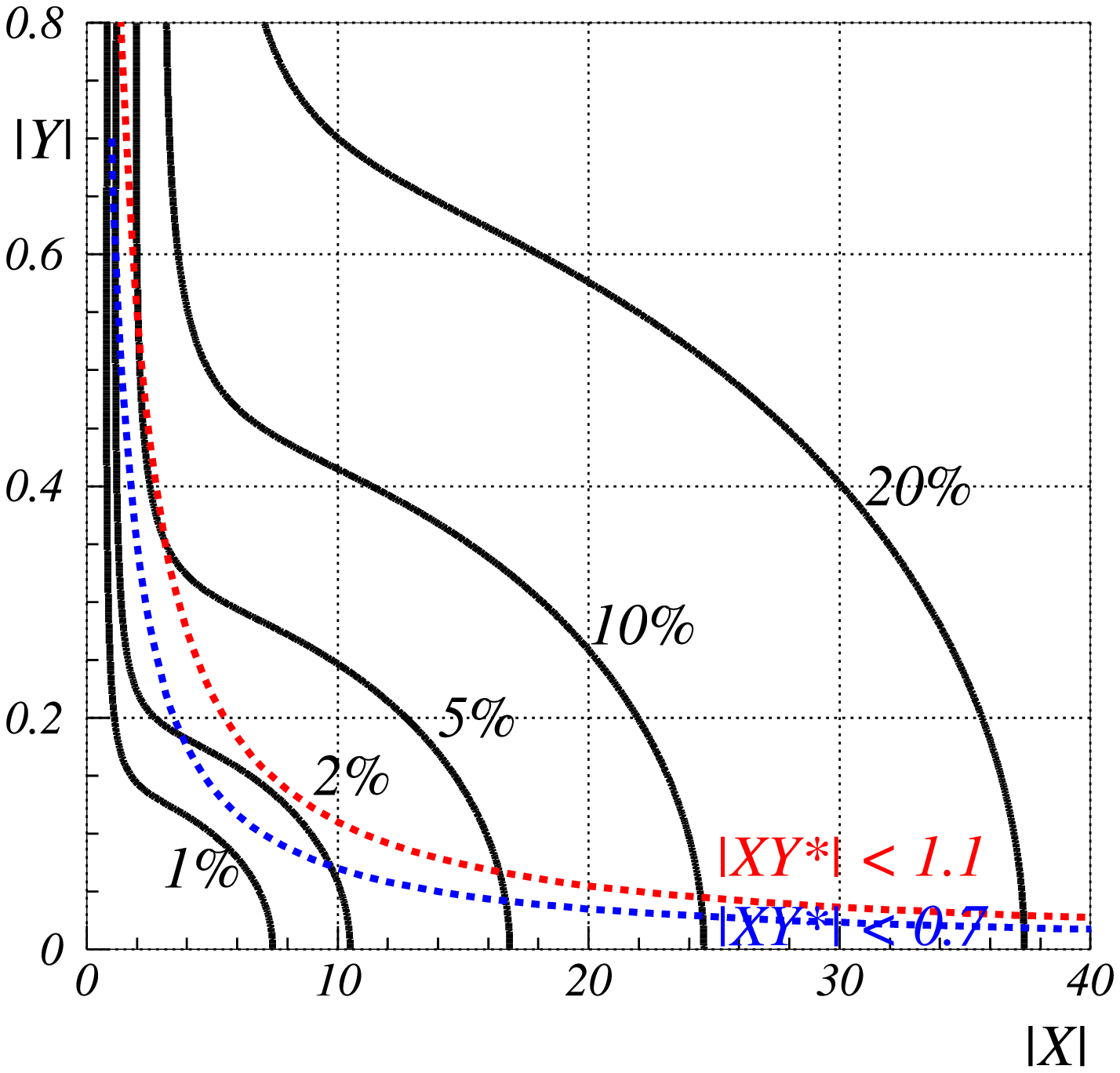}
\includegraphics[origin=c, angle=0, scale=0.5]{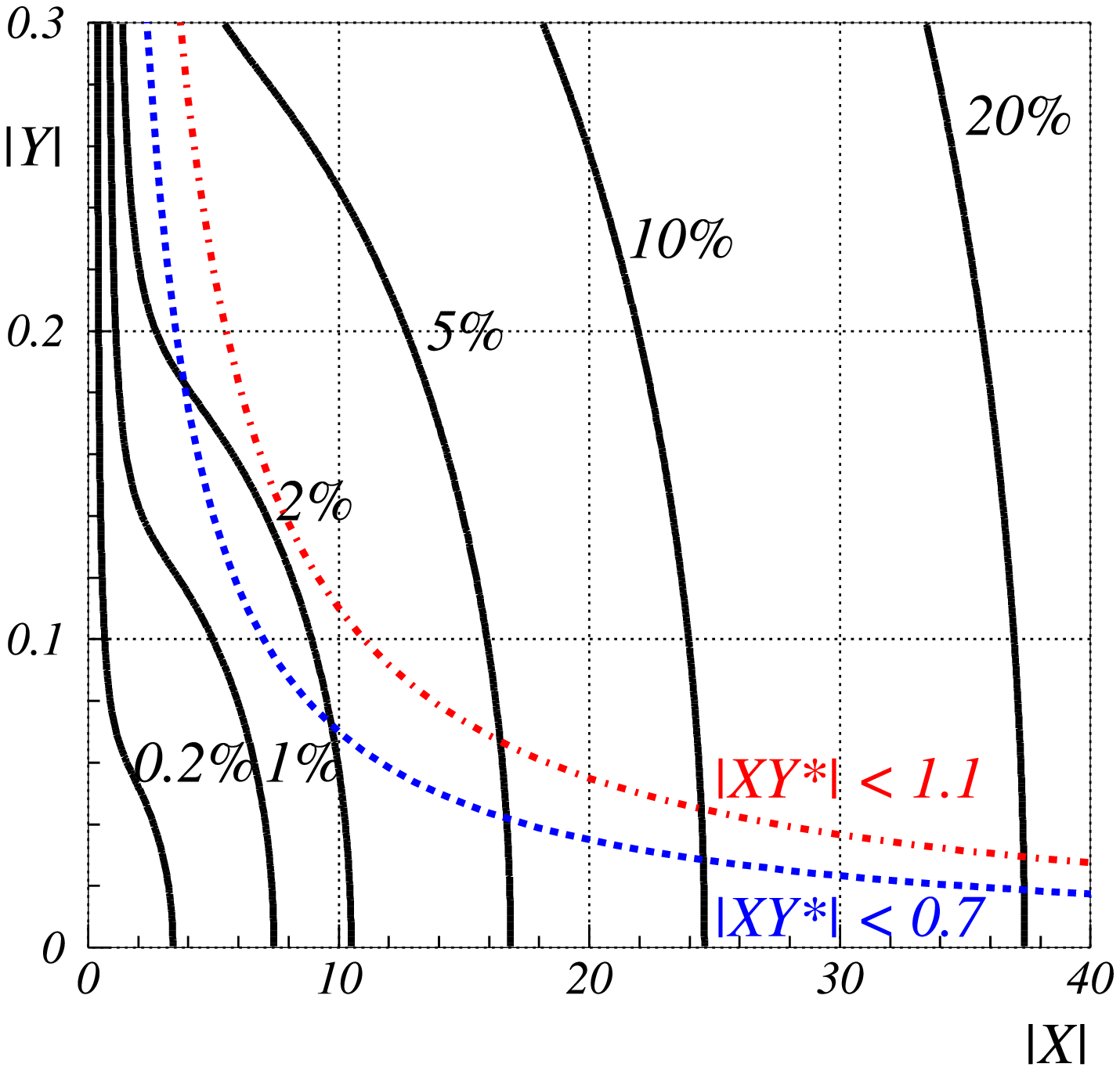}
%\vspace*{-5mm}
\caption{Contours of BR$(t\to H^\pm b)\times {\rm BR}(H^\pm\to cb$) in the plane [$X$, $Y$] with $|Z|=0.1$ for
$m_{H^\pm}=80$ GeV.
The constraint from $b\to s\gamma$ is shown as $|XY^*| < 1.1$
for Re$(XY^*) < 0$, and $|XY^*| < 0.7$ for Re$(XY^*) >0$.
We take $m_s(Q=m_{H^\pm})=0.055$ GeV and show the range $0 < |Y| < 0.8$  (left panel) and $0 < |Y| < 0.3$
(right panel). }
\label{fig:brtbcb80}
\end{center}
\end{figure}

\begin{figure}[t]
\begin{center}
\includegraphics[origin=c, angle=0, scale=0.5]{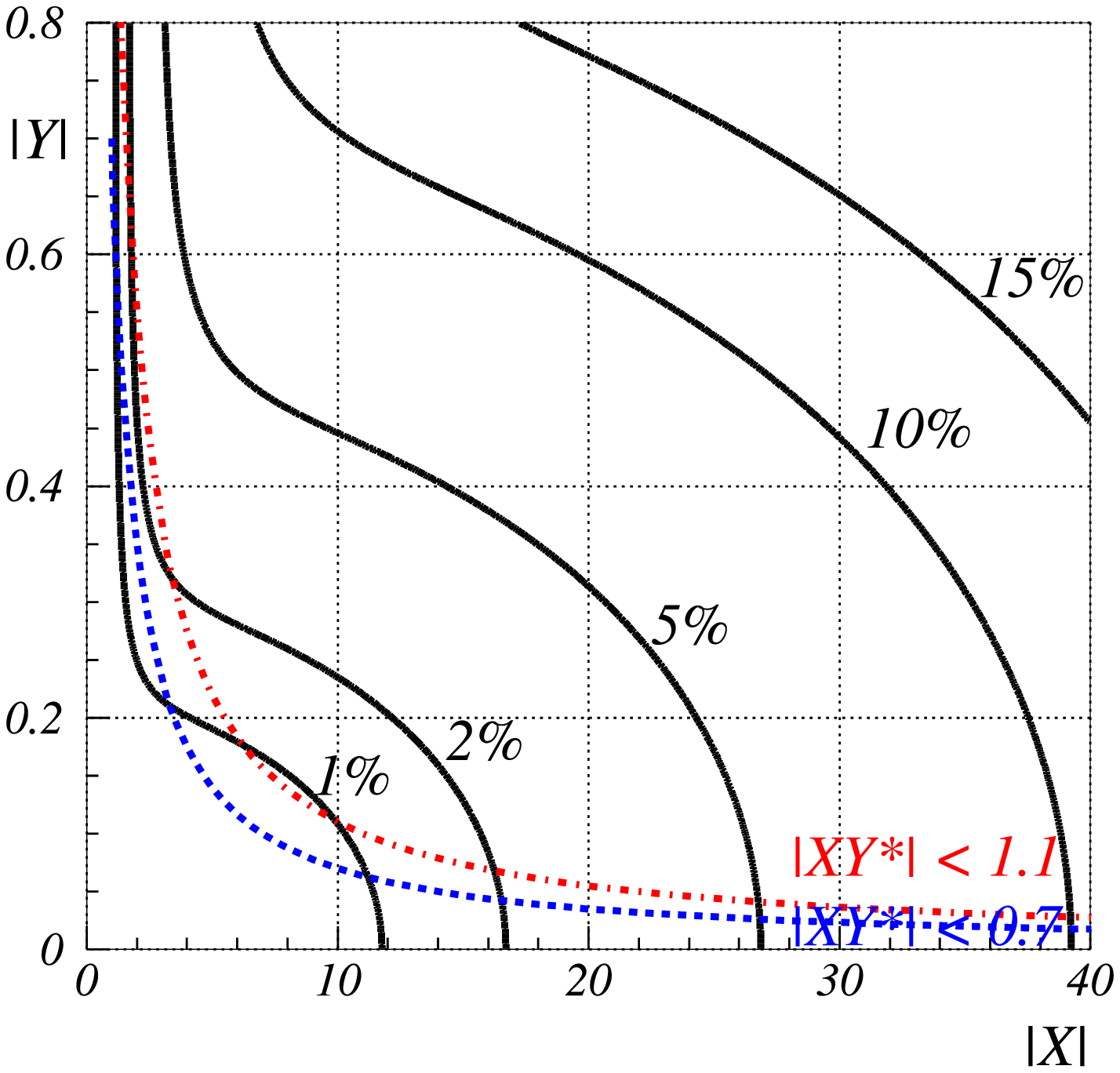}
\includegraphics[origin=c, angle=0, scale=0.5]{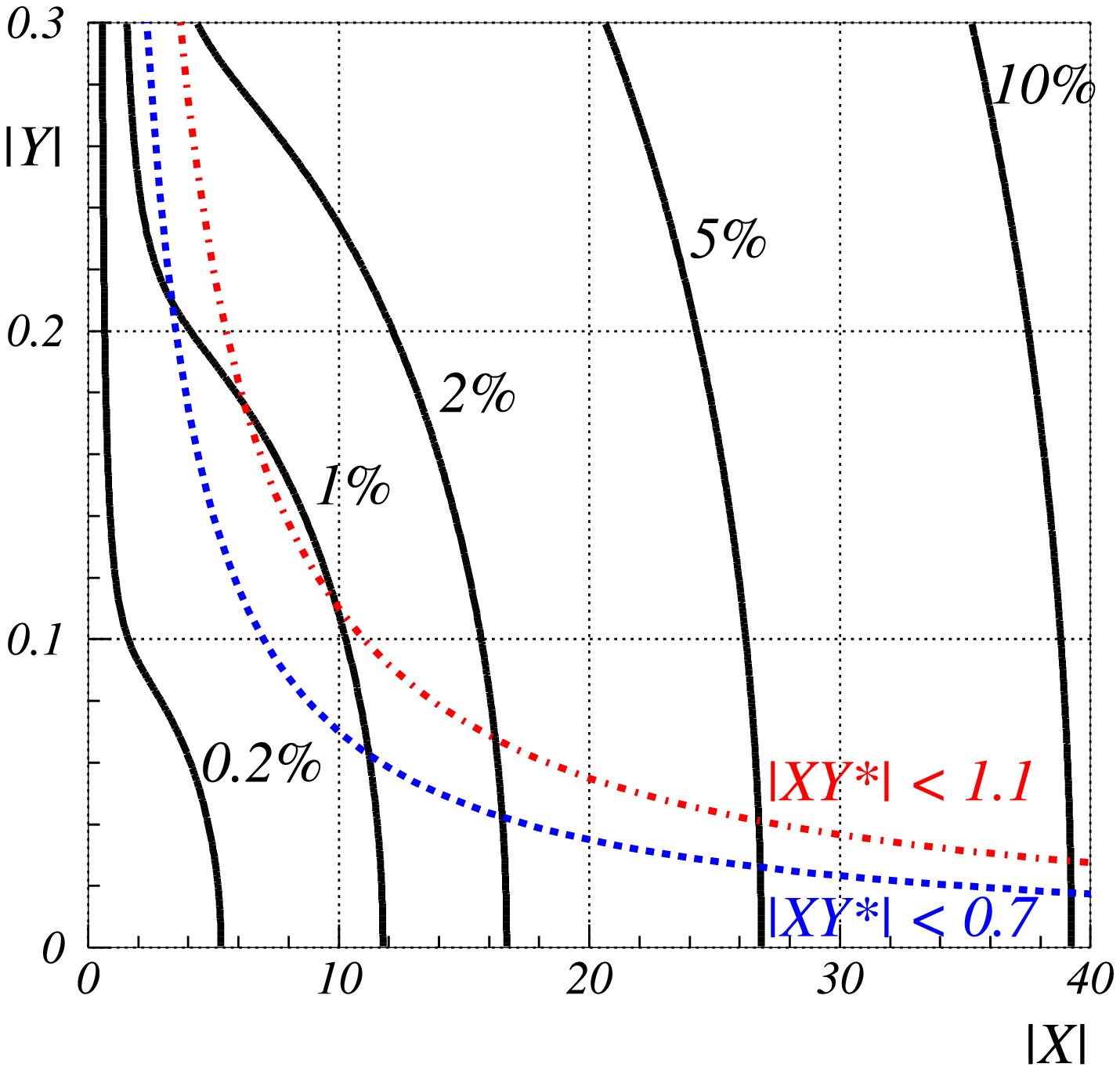}
%\vspace*{-5mm}
\caption{Contours of BR$(t\to H^\pm b)\times {\rm BR}(H^\pm\to cb$) in the plane [$X$, $Y$] with $|Z|=0.1$ for
$m_{H^\pm}=120$ GeV.
The constraint from $b\to s\gamma$ is shown as $|XY^*| < 1.1$
for Re$(XY^*) < 0$, and $|XY^*| < 0.7$ for Re$(XY^*) >0$.
We take $m_s(Q=m_{H^\pm})=0.055$ GeV and show the range $0 < |Y| < 0.8$  (left panel) and $0 < |Y| < 0.3$
(right panel). }
\label{fig:brtbcb120}
\end{center}
\end{figure}

Increased sensitivity to the plane of $[X,Y]$ can be achieved by requiring a $b$-tag on the jets 
which originate from the decay of $H^\pm$.
In Figs.~(\ref{fig:brtbcb80}) and (\ref{fig:brtbcb120}), for $m_{H^\pm}=80$ GeV and $m_{H^\pm}=120$ GeV
respectively, we show contours of
\begin{equation}
{\rm BR}(t\to H^\pm b)\times {\rm BR}(H^\pm\to cb) \,\, .
\label{cbcb}
\end{equation}

With the extra $b$-tag, as described in eq.~(\ref{gain_from_btag}), the sensitivity should reach  
${\rm BR}(t\to H^\pm b)\times {\rm BR}(H^\pm\to cb) > 0.5\%$, and perhaps as low 
as $0.2\%$. In the latter case, one can see from  Figs.~(\ref{fig:brtbcb80}b) and (\ref{fig:brtbcb120}b)
that a large part of the regions of $|X|<5$ (for $m_{H^\pm}=120$ GeV) and $|X|<3$ (for $m_{H^\pm}=80$ GeV) 
could be probed, even for $|Y|<0.2$. Therefore there would be sensitivity to a sizeable region of the 
parameter space of $[X,Y]$ which is not excluded
by $b\to s\gamma$, a result which is in contrast to the above case where no $b$-tag is applied
to the $b$-jets originating from $H^\pm$. We encourage a dedicated search for $t\to H^\pm b$
and $H^\pm \to cb$ by the Tevatron and LHC collaborations. Such a search
would be a well-motivated extension and application of the searches which have already been carried out in 
\cite{Aaltonen:2009ke} and
\cite{ATLAS:search}, and would offer the possibility of increased sensitivity to the fermionic couplings and
mass of $H^\pm$ in the A2HDM and a MHDM.

\section{Conclusions}

Light charged Higgs bosons ($H^\pm$) are being searched for in the decays of top quarks
($t\to H^\pm b$) at the Tevatron and at the LHC. Separate searches are being
carried out for the decay channels $H^\pm \to cs$ and 
$H^\pm \to \tau\nu$, with comparable sensitivity to the mass and fermionic couplings of $H^\pm$.  
The searches for $H^\pm \to cs$ in \cite{Aaltonen:2009ke} and
\cite{ATLAS:search} look for a peak at $m_{H^\pm}$ in the dijet invariant mass distribution, 
with the assumption that neither of the quarks is a $b$ quark.

In some models with two or more Higgs doublets (the Aligned 2HDM and
a MHDM with three or more scalar doublets) the branching ratio for
$H^\pm \to cb$ can be as large as $80\%$. Moreover, such a $H^\pm$ 
could be light enough to be
produced via $t\to H^\pm b$, as well as respect the stringent constraints from $b\to s\gamma$ on both $m_{H^\pm}$ and 
the fermionic couplings of $H^\pm$. This is in contrast to $H^\pm$ in other 2HDMs for which a large
branching ratio for $H^\pm \to cb$ is possible
(such as the flipped 2HDM for $m_{H^\pm} < m_t$), but one expects $m_{H^\pm} > m_t$ in order to comply with 
the measured value of $b\to s\gamma$.
In the context of the Aligned 2HDM and a MHDM 
we suggested that a dedicated search for $t\to H^\pm b$ and $H^\pm \to cb$ would
probe values of the fermionic couplings of $H^\pm$ which are currently not excluded by measurements of $b\to s\gamma$.
Such a search would require a $b$-tag of one of the jets originating from $H^\pm$, and would afford
sensitivity to a smaller value of the branching ratio of $t\to H^\pm b$ than that obtained in the ongoing searches, which 
currently do not make use of this additional $b$-tag.
We emphasised that a dedicated search for $t\to H^\pm b$ and $H^\pm \to cb$ at the Tevatron and LHC would be 
a well-motivated and (possibly) straightforward extension of the ongoing searches for $t\to H^\pm b$ with decay
$H^\pm \to cs$.

\section*{Acknowledgements}
A.G.A was supported by a Marie Curie 
Incoming International Fellowship, FP7-PEOPLE-2009-IIF, Contract No. 252263. J. H.-S acknowledges the financial 
support of SNI, PROMEP and VIEP-BUAP. S.M is supported in part by the NExT Institute.
J. H.-S. thanks the University of Southampton and the Rutherford Appleton Laboratory for hospitality
during his visit to the NExT Institute where part of this work was carried out.
Useful comments from A. Stuart and A. Lytle are gratefully acknowledged.

\end{document}